\documentclass[reprint,onecolumn,notitlepage,
showpacs,preprintnumbers,amssymb,nofootinbib,superscriptaddress,aps]{revtex4-1}

\usepackage{etoolbox}
\patchcmd{\section}
{\centering}
{\raggedright}
{}
{}
\patchcmd{\subsection}
{\centering}
{\raggedright}
{}
{}

\def\b{\beta}

\def\g{\gamma}

\def\l{\lambda}

\def\m{\mu}
\def\n{\nu}
\def\s{\sigma}

\def\b4{\xoverline{4}}

\def\b{\beta}

\def\tr{\text{tr}}

\usepackage{amssymb,amsmath,amsfonts,mathrsfs,enumerate,wrapfig}

\usepackage[margin=2cm]{geometry}
\usepackage[utf8]{inputenc}
\usepackage[T1]{fontenc}
\usepackage{lmodern}

\usepackage{mmacells}

\mmaDefineMathReplacement[≤]{<=}{\leq}
\mmaDefineMathReplacement[≥]{>=}{\geq}
\mmaDefineMathReplacement[≠]{!=}{\neq}
\mmaDefineMathReplacement[→]{->}{\to}[2]
\mmaDefineMathReplacement[⧴]{:>}{:\hspace{-.2em}\to}[2]
\mmaDefineMathReplacement{∉}{\notin}
\mmaDefineMathReplacement{∞}{\infty}
\mmaDefineMathReplacement{𝕕}{\mathbbm{d}}

\mmaSet{
	morefv={gobble=2},
	linklocaluri=mma/symbol/definition:#1,
	morecellgraphics={yoffset=1.9ex}
}

\usepackage{latexsym}
\usepackage{multirow}
\usepackage{textcomp}

\usepackage{subfig}
\usepackage{framed}

\usepackage[]{hyperref}

\usepackage{colortbl}

\makeatletter

\def\CT@@do@color{%
	\global\let\CT@do@color\relax
	\@tempdima\wd\z@
	\advance\@tempdima\@tempdimb
	\advance\@tempdima\@tempdimc
	\advance\@tempdimb\tabcolsep
	\advance\@tempdimc\tabcolsep
	\advance\@tempdima2\tabcolsep
	\kern-\@tempdimb
	\leaders\vrule
	\hskip\@tempdima\@plus  1fill
	\kern-\@tempdimc
	\hskip-\wd\z@ \@plus -1fill }
\makeatother

\usepackage{makecell}
\usepackage{color}
\usepackage{pifont}
\usepackage{appendix}
\usepackage{cancel}

\usepackage{import}
\usepackage{subfiles}
\usepackage{filecontents}
\usepackage{threeparttable,tablefootnote}
\usepackage{slashed}
\usepackage{mathtools}
\usepackage{pbox}

\definecolor{lightgray}{gray}{0.9}

\usepackage{fancyhdr, graphicx}

\begin{document}



\title{\texttt{\huge CoDEx} : Wilson coefficient calculator\\ connecting SMEFT to UV theory}

\author{Supratim Das Bakshi}
\email{sdbakshi13@gmail.com}
\affiliation{Department of Physics, Indian Institute of Technology, Kanpur-208016, India}

\author{Joydeep Chakrabortty}
\email{joydeep.chakrabortty@gmail.com}
\affiliation{Department of Physics, Indian Institute of Technology, Kanpur-208016, India}

\author{Sunando Kumar Patra}
\email{sunando.patra@gmail.com}
\affiliation{Department of Physics, Indian Institute of Technology, Kanpur-208016, India}

\begin{abstract}

	\mmaInlineCell{Input}{\mmaDef{CoDEx}} is a {\em Mathematica$^\text{\tiny \textregistered}$} package that calculates the Wilson Coefficients (WCs) corresponding to effective operators up to mass dimension-6. Once the part of the Lagrangian involving single as well as multiple degenerate heavy fields, belonging to some Beyond Standard Model (BSM) theory, is given, the package can then integrate out propagators from the tree as well as 1-loop diagrams of  that BSM theory. It then computes the associated WCs up to 1-loop level, for two different bases: \mmaInlineCell{Code}{"Warsaw"} and \mmaInlineCell{Code}{"SILH"}. \mmaInlineCell{Code}{\mmaDef{CoDEx}} requires only very basic information about the heavy field(s), e.g., Colour, Isospin, Hyper-charge, Mass, and Spin. The package first calculates the WCs at the high scale (mass of the heavy field(s)). We then have an option to perform the renormalisation group evolutions (RGEs) of these operators in \mmaInlineCell{Code}{"Warsaw"} basis, a complete one (unlike \mmaInlineCell{Code}{"SILH"}), using the anomalous dimension matrix. Thus, one can get all effective operators at the electro-weak scale, generated from any such BSM theory, containing heavy fields of spin: 0, 1/2, and 1. We have provided many example models (both here and in the package-documentation) that more or less encompass different choices of heavy fields and interactions. Relying on the status of the present day precision data, we restrict ourselves up to dimension-6 effective operators. This will be generalised for any dimensional operators in a later version.

\end{abstract}

\maketitle
\thispagestyle{fancy}

%


%
%

\begin{small}
\noindent
{\bf Program Summary}							\\

{\em Program Title: \texttt{CoDEx}}                                          \\
{\em Version: \texttt{1.0.0}}					\\
{\em Licensing provisions: \texttt{CC By 4.0}}                         \\
{\em Programming language: \texttt{Wolfram Language$^\text{\tiny \textregistered}$}}                                   \\
{\em URL:} \url{https://effexteam.github.io/CoDEx}					\\
{\em Send BUG reports and Questions:} \href{mailto:effex.package@gmail.com}{\texttt{effex.package@gmail.com}}					\\
\end{small}

\newpage

\section{Introduction}
It is a perplexing time for particle physics. On one side we are cherishing the discovery of the Standard Model (SM)-Higgs like particle, considered to be the pinnacle of success of the SM, on the other hand we have enough reason to believe  the existence of theories beyond it (BSM). To address the shortcomings of the SM, many BSM scenarios are proposed at very different scales. It is believed that any such theory, which contains the SM as a part of it, will affect the electro-weak and the Higgs sector. Thus the precision observables are expected to carry the footprints of the $new$ physics, unless it is in the decoupling limit.

The ongoing and proposed future experiments are expected to improve the sensitivity of these precision observables at per mille level. Thus we can indirectly estimate the allowed room left for some BSM physics, even in the case of non-observation of new resonances. This motivates us to look into the BSM scenario through the tinted glass of Standard Model effective field theory (SMEFT). The basic idea of SMEFT is quite straightforward: integrate out heavy non-SM degrees of freedom and capture their impact through the higher mass dimensional operators -- $\sum_{i}(1/\Lambda^{d_i - 4}) C_i \mathcal{O}_i$. Here, $d_i$ is the mass dimensionality of the operator $\mathcal{O}_i$ (starts from $5$), and $C_i$ is the corresponding Wilson coefficient -- function of BSM parameters. It is important to note that the choice of operator basis, i.e., explicit structure of $\mathcal{O}_i$'s is not unique.  Among different choices we restrict us to \mmaInlineCell{Code}{"SILH"} \cite{Giudice:2007fh,Contino:2013kra} and \mmaInlineCell{Code}{"Warsaw"} \cite{Jenkins:2013zja, Jenkins:2013wua,Alonso:2013hga} bases. These bases can be transformed from one to another
. $\Lambda$ is the cut off scale at which all WCs are computed ($C_i(\Lambda)$), and usually identified as the mass of the heavy field being integrated out. This EFT approach relies on the validity of the perturbative expansion of the S-matrix in the powers of $\Lambda^{-1}$ (UV-scale), and the resultant series is expected to pass the convergence test. As this scale is higher than the scale $M_Z$, where the precision test is performed, dimension-6 operators are more suppressed than the dimension-5 ones and so on. Now, where to truncate the $1/\Lambda$ series? This decision is made case by case, based on the achieved(expected) precision level of the observables at present(future) experiments\footnote{For further interested readers, see \cite{Furnstahl:2015rha}.}.  One can consult these lectures  \cite{Georgi:1994qn,Kaplan:1995uv,Manohar:1996cq,Burgess:2007pt,Rothstein:2003mp} where effective field theory has been introduced and discussed in great detail. Several other packages and libraries are available in the literature, which do various things regarding SMEFT operators and the corresponding Wilson Coefficients, from basis transformation to running of the coefficients \cite{Gripaios:2018zrz,Falkowski:2015wza,Celis:2017hod,Criado:2017khh,Aebischer:2018bkb}.

Now the nagging questions are: ({\em a}) Why use SMEFT instead of doing the full calculation, using the supposedly more accurate BSM Lagrangian? ({\em b}) How can one ensure that the difference between the results, computed in SMEFT approach using a truncated S-matrix and those obtained using the full BSM theory, is imperceptible (in the precision tests)?

The computation with the full BSM is involved and tedious, and that too at loop level. The cut-off $\Lambda$ is chosen in such a way that the $M_Z/\Lambda$ series is converging, which ensures that the truncation of this series at some finite order is safe and sufficient. Even then, the question remains: how do we connect the physics of two different scales, namely UV and the $M_Z$? The WCs, that we are computing using SMEFT, are at the scale $\Lambda$, but the observables are measured at $M_Z$ scale. Hence, we need to evolve the $C_i(\Lambda)$ to obtain $C_i(M_Z)$, using the anomalous dimension matrix $(\gamma)$.  While performing the renormalisation group evolutions (RGEs) of the $C_i'$s, we need to choose the $\gamma$ carefully, as it is basis dependent. Thus we need to choose only those bases, in which the precision observables are defined, and it is important to ensure that the basis we are working with is a complete one. As the matrix $\gamma$ contains non-zero off-diagonal elements, it is indeed possible to generate, through RGEs, some new effective operators, which were absent at $\Lambda$ scale. These effective field theory approach has been successfully used in the context of precision data and Higgs phenomenology, for details see \cite{Henning:2014wua,Henning:2016lyp,Henning:2015alf,Elias-Miro:2013mua,Elias-Miro:2013eta,Jenkins:2013fya,Manohar:2013rga,Han:2004az,Cacciapaglia:2006pk,Bonnet:2011yx,Bonnet:2012nm,delAguila:2011zs,Grojean:2013kd,Contino:2016jqw,Brehmer:2015rna,Berthier:2015gja,Gorbahn:2015gxa,Berthier:2015oma,Skiba:2010xn,Khandker:2012zu,Englert:2013wga,Banerjee:2012xc,Banerjee:2013apa}.

With this backdrop, we introduce \mmaInlineCell{Code}{\mmaDef{CoDEx}}, a {\em Mathematica$^\text{\tiny \textregistered}$} package, which can integrate out the heavy field propagator(s) from tree as well as 1-loop processes and can generate SM effective operators up to dimension-6. It also provides the Wilson Coefficients as a function of BSM parameters. The draft is organised in the following manner: in section \ref{sec:theo}, we have briefly discussed the underlying principle of \mmaInlineCell{Code}{\mmaDef{CoDEx}} from a theoretical perspective. Details about downloading and installation are in section \ref{sec:install}. In the remainder of section \ref{sec:pack}, we provide a guideline to define the heavy field(s) and build the working Lagrangian (section \ref{sec:build}), provide a list of all the functions that are necessary to run \mmaInlineCell{Code}{\mmaDef{CoDEx}} in detail (section \ref{sec:howtorun}), and explain the way \mmaInlineCell{Code}{\mmaDef{CoDEx}} takes care of the RGEs of the WCs down to the electro-weak scale (section \ref{sec:rgflow}). In the next section, we provide the user with one detailed work-flow to use the package for a model with a single heavy electro-weak $SU(2)_L$ real singlet scalar in it (section \ref{sec:detailex}). Sequentially following these steps should enable one to find out the effective operators up to mass dimension-6 and the respective WCs at the high scale for that model. In the appendix  we have provided various example models to encapsulate different types of fields that are used frequently to build BSM scenarios. One can consult the Refs. \cite{Jenkins:2013zja, Jenkins:2013wua,Alonso:2013hga,Wells:2015cre} regarding the running of the SM effective operators.

\section{The underlying Principle}\label{sec:theo}

In this section, we will briefly discuss the adopted method, based on the idea of \textit{Co}variant \textit{D}erivative \textit{Ex}pansion (CDE), to integrate out the heavy fields to compute the Wilson Coefficients (WCs). This was introduced in \cite{Gaillard:1985uh}, and then extended in \cite{Cheyette:1985ue}. As we are performing this {\it ``integrating out of the field(s)"} order by order in perturbation theory, we need to respect the gauge invariance at each and every step. The perturbative expansion thus demands to be done in terms of some gauge covariant quantities. Thus covariant derivative is the `chosen one'. CDE is not only restricted to quantify the integrating out of heavy fields, rather has a wider impact; see \cite{Henning:2014wua,Henning:2016lyp,Henning:2015alf} for detail.   
The method of integrating out different types of heavy fields using functional methods and the basis dependency are discussed in many places in the literature, see \cite{Lehman:2015via,Chiang:2015ura,Huo:2015exa,Huo:2015nka,Lehman:2015coa,Wells:2015uba,Drozd:2015rsp,delAguila:2016zcb,Fuentes-Martin:2016uol,Ellis:2016enq, Falkowski:2015wza} for detail.

Considering the status of present and prospect of future experiments, we can adjudge ourselves safe, when we restrict ourselves to only dimension-6 operators including the tree and 1-loop parts of the effective action. The modus operandi of \mmaInlineCell{Code}{\mmaDef{CoDEx}} is based on the method of CDE discussed in \cite{Henning:2014wua,Henning:2016lyp,Henning:2015alf}. 
Briefly, here is what \mmaInlineCell{Code}{\mmaDef{CoDEx}} can give you:
\begin{itemize}
	\item It will integrate out the heavy BSM field(s) while respecting the gauge invariance.
	\item It will generate Wilson Coefficients at tree and(or) 1-loop level.
	\item It will perform RG evolutions of the effective operators generated at some high scale and provide the operators at the electro-weak scale.
\end{itemize}

Let us consider a BSM Lagrangian in the following form
\begin{equation}
\mathcal{L}^{(BSM)} \equiv \mathcal{L}^{(BSM)} ( \phi , \Phi )=\mathcal{L}^{(\Phi )} + \mathcal{L}^{(\phi )} + \mathcal{L}^{(\phi , \Phi )}_{int}\,,
\end{equation}
where $\phi$ and $\Phi$ represent light (SM) and heavy (BSM) fields respectively.  Here, $\mathcal{L}^{(\phi )}, \mathcal{L}^{(\Phi )}$, and $\mathcal{L}^{(\phi , \Phi )}_{int}$ are three different sectors of the BSM Lagrangian containing respectively only-heavy fields, only-light fields, and heavy-light both.

To proceed further, we have to first solve the following  Euler-Lagrange (EL) equation to compute the solution for the heavy field:
\begin{equation}
\frac{\partial}{\partial \Phi} \mathcal{L}^{(BSM)}(\phi , \Phi ) = \mathcal{D}_{\mu} \ \frac{\partial}{\partial (\mathcal{D}_{\mu} \Phi)} \mathcal{L}^{(BSM)}(\phi , \Phi ) ,
\label{el0}
\end{equation}
where $\mathcal{D}_{\mu}$ is the covariant derivative corresponding the heavy field $\Phi$. To find out the EL-equations, we will concentrate on the $\mathcal{L}^{(BSM)}( \phi , \Phi ) $ part only. If we express this part of the Lagrangian as a polynomial in $\Phi$, e.g.,  $\mathcal{L}^{(\phi)}_{I} . \ \Phi + \mathcal{L}^{(\phi)}_{II} . \ (\Phi)^{2}  + \ldots$, the coefficients can then be written as:
\begin{equation}
\mathcal{L}^{(\phi)}_{I} = \ \mathcal{O}_{\mathcal{D}} \ . \ \hat{\Phi} ,
\label{od0}
\end{equation} 
where $\mathcal{O}_{\mathcal{D}}$ contains the information regarding the covariant derivative of the heavy field and the functional of the light (SM) fields. In general, $\mathcal{O}_{\mathcal{D}}$ is in the form of an elliptic operator, e.g. $\mathcal{O}_{\mathcal{D}} = \mathcal{D}^{2} + M^{2} + \mathcal{L}_{II}^{\phi} $, where $M$ is the mass of heavy field to be integrated out. Thus the heavy field solution can be rewritten as:
\begin{equation}
\hat{\Phi} = [\mathcal{O}_{\mathcal{D}}]^{-1} \ . \ \mathcal{L}^{(\phi)}_{I} .
\end{equation}
The operator $[\mathcal{O}_{\mathcal{D}}]^{-1}$ can be Taylor-expanded, where terms are suppressed by $M^{2n}$ ($n$  takes integer values starting from 1). This series is convergent as $M$ is much larger than the allowed maximum momentum transfer in the low energy theory. Thus we can truncate this series based on our requirement of the mass dimension of the effective operator. It is important to note that in the theories where the $\mathcal{L}^{(\phi)}_{I}$ term is absent,  any tree-level effective operator will not be generated after integrating out $\Phi$. 

The next task is to compute these effective operators at loop level. We will restrict our computation up to 1-loop, relying on the precision of present data. The effective Lagrangian at 1-loop level is given by \cite{Henning:2014wua,Henning:2016lyp,Henning:2015alf}:
{\small
\begin{align}\label{leff1loop}
\mathcal{L}^{(dim-6)}_{1-loop}[\phi] =&\frac{c_s}{(4\pi)^2} \, \tr \, \Bigg\{ \frac{1}{m^2}\Bigg[ -\frac{1}{60} \, \big(P_{\m}G_{\m\n}'\big)^2 - \frac{1}{90} \, G_{\m\n}'G_{\n\s}'G_{\s\m}' -\frac{1}{12} \, (P_{\m}U)^2 
- \frac{1}{6}\, U^3 - \frac{1}{12}\, U G_{\m\n}'G_{\m\n}' \Bigg] \nonumber \\ 
&+ \frac{1}{m^4} \Bigg[\frac{1}{24} \, U^4 + \frac{1}{12}\, U \big(P_{\m}U\big)^2 + \frac{1}{120}\, \big(P^2U\big)^2 +\frac{1}{24} \, \Big( U^2 G'_{\m\n}G'_{\m\n} \Big) 
- \frac{1}{120} \, \big[(P_{\m}U),(P_{\n}U)\big] G'_{\m\n} \nonumber \\ 
& - \frac{1}{120}\, \big[U[U,G'_{\m\n}]\big] G'_{\m\n} \Bigg] + \frac{1}{m^6} \Bigg[-\frac{1}{60} \, U^5 - \frac{1}{20} \, U^2\big(P_{\m}U\big)^2 - \frac{1}{30} \, \big(UP_{\m}U\big)^2 \Bigg] 
+ \frac{1}{m^8} \bigg[ \frac{1}{120} \, U^6 \bigg]\Bigg\} \,.
\end{align}
}
where $c_{s} = \frac{1}{2}, 1, - \frac{1}{2}$, and $\frac{1}{2}$ for real scalar, complex scalar, fermion, and gauge boson respectively. Here, $P_{\mu} = i \ \mathcal{D}_{\mu} \ \text{and} \ G'_{\mu \nu}=[\mathcal{D}_{\mu} , \mathcal{D}_{\nu}]$. $U$ is a collection of coefficients of the terms which are bi-linear in the heavy field. $U$ can be written in matrix form of 
$\delta\Big[ \mathcal{L}^{(BSM)}(\phi , \Phi )\Big]/\delta \Phi_i \delta \Phi_j$, evaluated at $\hat{\Phi}_i$ \cite{Henning:2014wua,Henning:2016lyp,Henning:2015alf}. It is important to note down that `$\tr$' in the above equation is the trace performed over the internal symmetry indices.

Once we find the effective operators and their respective Wilson Coefficients at high scale, i.e., scale of new physics, we can run them down to the electro-weak scale. These operators are evolved with energy scale as:
\begin{equation}
\frac{d \mathcal{O}_i}{d\ln\mu}=\gamma_{ij} \mathcal{O}_j,
\end{equation}
where $\gamma_{ij}$ is the anomalous dimension matrix. This is also implemented in \mmaInlineCell{Code}{\mmaDef{CoDEx}}, but only for \mmaInlineCell{Code}{"Warsaw"} basis, as it is the complete one.

\section{The package, in detail}\label{sec:pack}

\subsection{Installing and Loading {\large \texttt{CoDEx}}}\label{sec:install}

Installing \mmaInlineCell{Input}{\mmaDef{CoDEx}} is quite straightforward. This can be done in one of the following ways:

\subsubsection{Automatic Installation}
\mmaInlineCell{Input}{\mmaDef{CoDEx}} can be installed in a Mathematica environment by downloading and importing the installer file `\texttt{install.m}' available at \href{https://github.com/effExTeam/CoDEx-1.0.0/raw/master/install.m}{this link}. 
The installer can also be automatically loaded inside Mathematica environment by using the command\footnote{This requires a live internet connection and Mathematica should be able to connect to the internet.}
\begin{mmaCell}{Code}
		Import["https://github.com/effExTeam/CoDEx-1.0.0/raw/master/install.m"];
\end{mmaCell}
This loads two functions in the working kernel: 
`\mmaInlineCell{Code}{InstallCoDEx}' and `\mmaInlineCell{Code}{InstallCoDExQuiet}'. Typical way of running them is:
\begin{mmaCell}{Input}
		\mmaDef{InstallCoDEx}[]
\end{mmaCell}

One can use \mmaInlineCell{Code}{InstallCoDExQuiet} instead of \mmaInlineCell{Code}{InstallCoDEx}, which is equivalent to running:
\begin{mmaCell}[index=2]{Input}
		\mmaDef{InstallCoDEx}[AutoDisableInsufficientVersionWarning\(\pmb{\to}\)True,
					AutoOverwriteCoDExDirectory\(\pmb{\to}\)True]
\end{mmaCell}

\begin{table}[ht]
	\caption{Running `InstallCoDEx' can be customized using its options:}
	\begin{center}
	\renewcommand*{\arraystretch}{2}
	\begin{tabular}{|cll|}
		\hline
		\rowcolor[gray]{.92}
		\textit{Options}$~~~$ & \textit{Default}$~~~$ & \textit{Detail} \\
		\hline
		\rowcolor[gray]{.92}
		$~~~$\mmaInlineCell{Code}{AutoDisableInsufficient-}$~~$ & \mmaInlineCell{Code}{None} & True $\to $ warning messages for notebooks created with$~~~$ \\
		\rowcolor[gray]{.92}
		\mmaInlineCell{Code}{VersionWarning} & & newer Mathematica version automatically disabled.\footnote{Needed to generate documentation in older versions. } \\
		\rowcolor[gray]{.92}
		& &	None$\to $ user will be asked by a dialog.\\
		\rowcolor[gray]{.92}
		& &	False$\to $ warnings will not be disabled. \\
		\rowcolor[gray]{.92}
		\mmaInlineCell{Code}{AutoOverwriteCoDEx-} & \mmaInlineCell{Code}{None} & True$\to $ Previous installations automatically deleted.\\
		\rowcolor[gray]{.92}
		\mmaInlineCell{Code}{\mmaUnd{Directory}} & & 		None$\to $ user will be asked by a dialog.\\
		\rowcolor[gray]{.92}
		& &		False$\to $ the directory will be overwritten. \\
		\rowcolor[gray]{.92}
		\mmaInlineCell{Code}{InstallCoDEx-} & \mmaInlineCell{Code}{False} & True$\to $ Install latest {\em development} version.\\
		\rowcolor[gray]{.92}
		\mmaInlineCell{Code}{\mmaUnd{DevelopmentVersion}} & &		False$\to $ Install the latest {\em stable} version. \\
		\rowcolor[gray]{.92}
		\mmaInlineCell{Code}{InstallCoDExTo} & \mmaInlineCell{Code}{"path"} & Specifies custom full path to installation directory. \\
		\hline
	\end{tabular}
	\end{center}
\end{table}

\subsubsection{Download archive file}

There is a provision to download and install the program locally.\\ 
The \mmaInlineCell{Input}{\mmaDef{CoDEx}} package is available in both \texttt{.zip} and \texttt{.tar} format in its `Github' repository. While using the `\mmaInlineCell{Code}{Install}' option from the Notebook menu inside Mathematica or manually extracting the downloaded archive file in the `\texttt{Applications}' folder inside the \mmaInlineCell{Code}{$UserBaseDirectory} works perfectly, installing \texttt{CoDEx} is made a lot easier by downloading and importing the installer file `\texttt{install.m}' available \href{https://github.com/effExTeam/CoDEx-1.0.0/raw/master/install.m}{here}.

If for some reason you choose to download the archive files and install CoDEx from them, this can still be done, using \mmaInlineCell{Code}{InstallCoDEx}. You just need to run it in a slightly different way.

You need to copy the path of the downloaded archive (e.g. if you are working in Windows and have downloaded the \texttt{.zip} file to the `Downloads' folder, your path will be \mmaInlineCell{Code}{"C:\...\Downloads\CoDEx.zip"}. Let's call it \mmaInlineCell{Code}{"path"} from now). 

Then run this command in your notebook after importing the `install.m' file:
\begin{mmaCell}[index=1]{Code}
		$PathToCoDExArchive = "path";
		\mmaDef{InstallCoDEx}[]
\end{mmaCell}

This goes through exactly the same steps as in the previous section, but instead of downloading the archive from the server, uses the local file.
After this, the package can always be loaded in {\em Mathematica} using:
\begin{mmaCell}{Input}
		Needs["CoDEx`"]
\end{mmaCell}

The installer file is not our creation. We have edited and simplified the installer for {\em FeynRules} \cite{Alloul:2013bka}. 

\begin{table}[ht]
	\caption{Main functions provided by \texttt{CoDEx}}
	\begin{center}
		\renewcommand*{\arraystretch}{2}
		\begin{tabular}{{|cl|}}
			\hline
			\rowcolor[gray]{.92}
			\textit{Function} &  \textit{Details}  \\
			\hline
			\rowcolor[gray]{.92}
			\mmaInlineCell{Code}{CoDExHelp}  &  Opens the \mmaInlineCell{Code}{\mmaDef{CoDEx}} guide, with all help files listed.\\ 
			\rowcolor[gray]{.92}
			\mmaInlineCell{Code}{treeOutput}  &  Calculates WCs generated from tree level processes.\\ 
			\rowcolor[gray]{.92}
			\mmaInlineCell{Code}{loopOutput}  &  Calculates WCs generated from 1-loop processes.\\ 
			\rowcolor[gray]{.92}
			\mmaInlineCell{Code}{codexOutput}  &  Generic function for WCs calculation with\\ 
			\rowcolor[gray]{.92}
			& choices for level, bases etc. given with \mmaInlineCell{Code}{\mmaDef{OptionValue}}s.\\
			\rowcolor[gray]{.92}
			\mmaInlineCell{Code}{defineHeavyFields}  &  Creates representation of heavy fields.\\ 
			\rowcolor[gray]{.92}
			&  Use the output to construct BSM Lagrangian.\\
			\rowcolor[gray]{.92}
			\mmaInlineCell{Code}{texTable}  &  Given a \mmaInlineCell{Code}{\mmaDef{List}}, returns the \LaTeX~output of a tabular\\ 
			\rowcolor[gray]{.92}
			& environment, displayed and/or copied to clipboard.\footnote{This is a simplified version of the package titled \texttt{TeXTableForm} \cite{TeXTableForm}} \\
			\rowcolor[gray]{.92}
			\mmaInlineCell{Code}{formPick}  &  Applied on a list of WCs from a specific operator basis,\\ 
			\rowcolor[gray]{.92}
			& reformats the output in the specified style. \\
			\rowcolor[gray]{.92}
			\mmaInlineCell{Code}{RGFlow}  &  RG Flow of WCs of dim. 6 operators in \mmaInlineCell{Code}{"Warsaw"} basis, \\
			\rowcolor[gray]{.92}
			&  from matching scale to a lower (arbitrary) scale.  \\
			\rowcolor[gray]{.92}
			\mmaInlineCell{Code}{initializeLoop}  &  Prepares the Isospin and Color symmetry generators \\
			\rowcolor[gray]{.92}
			&  for a specific model with a specific heavy field content. \\
			\rowcolor[gray]{.92}
			&  \mmaInlineCell{Code}{loopOutput} can only be run after this step is done.  \\
			\hline
		\end{tabular}
	\end{center}
\end{table}

\begin{figure}[ht]
	\centering
	\includegraphics[scale=0.2]{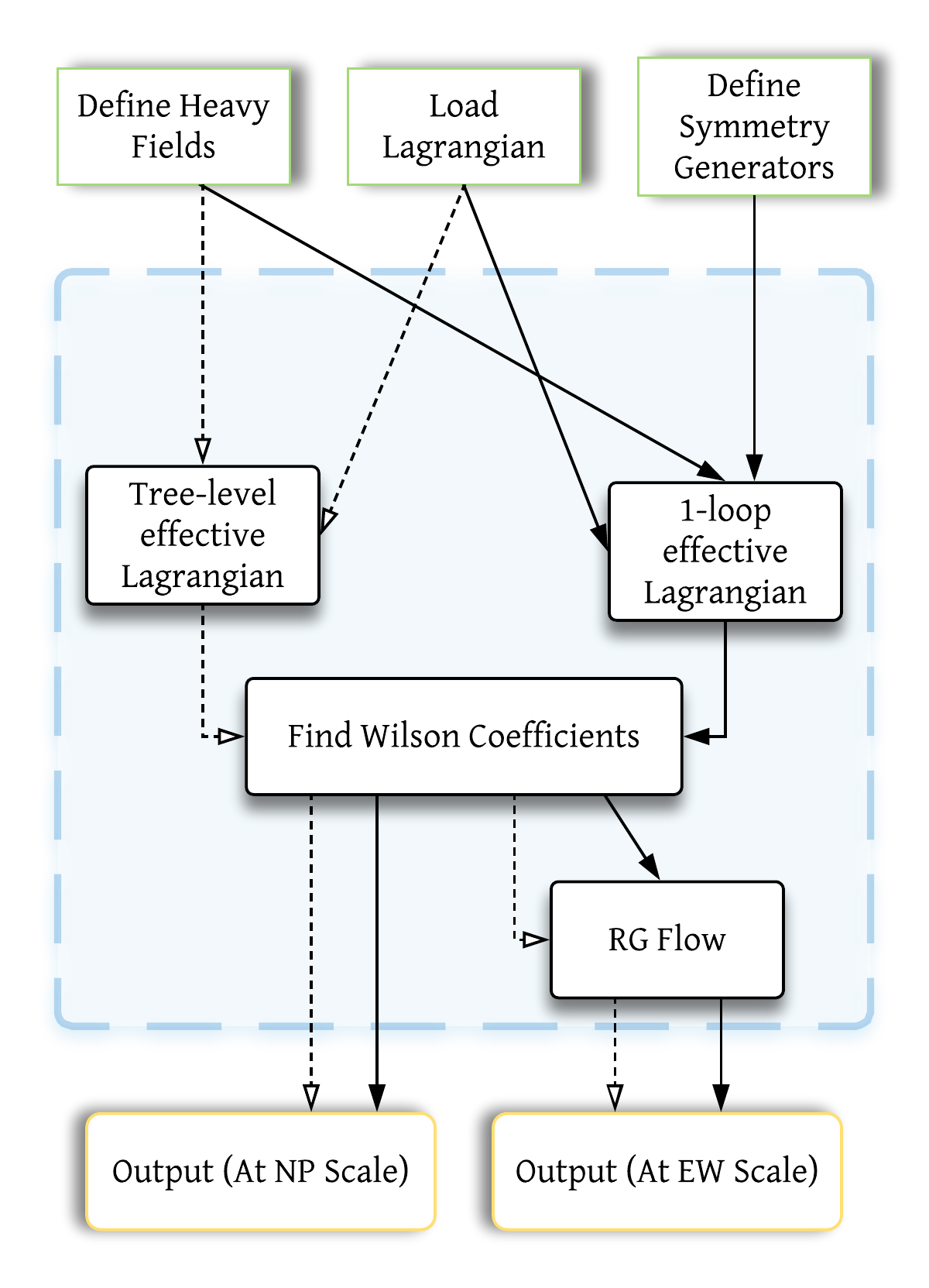}
	\caption{Flow-chart for \texttt{CoDEx}}
	\label{flowch}
\end{figure}

\subsection{How to build the Lagrangian}\label{sec:build}

\subsubsection{Building the Lagrangian: An example}\label{sec:buildEx}

Let us demonstrate this with a toy example where the Lagrangian is given in its traditional form:
\begin{equation}\label{lag1}
L(\Phi ,H)\supset -\eta~  \Phi ^a \Phi ^a \left| H\right| ^2 + 2 \kappa~ H \tau ^a \Phi ^a H^{\dagger }-\frac{\lambda _{\Phi }}{4} \left(\Phi ^a \Phi ^a\right)^2 \,.
\end{equation}
 Here, $\Phi$ is the heavy field which is going to be integrated out. 
 From the user-end, the information for this heavy field is fed in the code as:
 \begin{mmaCell}[label={}]{Code}
		fields =
		{ 
			{\mmaPat{fieldName}, \mmaPat{components}, \mmaPat{colorDim}, \mmaPat{isoDim}, \mmaPat{hyperCharge}, \mmaPat{spin}, \mmaPat{mass}}
		};
 \end{mmaCell}
which contains the required information (within \{\ldots\}) about the field to compute the WCs. The properties necessary to define the heavy field are listed in table \ref{tab:fieldForm}.
Now, this code is equally applicable for multiple heavy BSM fields. In this case, the field-definitions will be listed sequentially under the first set of curly braces (see section \ref{sec:buildheavy}). In the case of our example model, we have only one heavy field, a real triplet scalar (i.e. \mmaInlineCell{Code}{components} $\rightarrow$ 3, \mmaInlineCell{Code}{colorDim} $\rightarrow$ 1, \mmaInlineCell{Code}{isoDim} $\rightarrow$ 3, \mmaInlineCell{Code}{hypercharge} $\rightarrow$ 0, \mmaInlineCell{Code}{spin} $\rightarrow$ 0). Let us denote \mmaInlineCell{Code}{fieldName} $\rightarrow$ `ph' and \mmaInlineCell{Code}{mass} $\rightarrow$ `m'.
This represents the field content of our model in the correct way:
\begin{mmaCell}[index=1]{Code}
		fields =
		{
			{ph, 3, 1, 3, 0, 0, m}
		};
\end{mmaCell}


\begin{table}[ht]
	\caption{Generic form of individual fields:}\label{tab:fieldForm}
	\centering
	\renewcommand*{\arraystretch}{2}
	\begin{tabular}{|cl|}
		\rowcolor[gray]{.92}
		\hline
		\textit{Pattern} &  \textit{Details}  \\
		\hline
		\rowcolor[gray]{.92}
		\mmaInlineCell{Code}{fieldName} & field-name, used as array-head to construct the Lagrangian. \\
		\rowcolor[gray]{.92}
		\mmaInlineCell{Code}{components} & \text{number of components in the heavy field.} \\
		\rowcolor[gray]{.92}
		\mmaInlineCell{Code}{colorDim} & \text{dimensionality of the heavy field under } $SU(3)_C$ \\
		\rowcolor[gray]{.92}
		\mmaInlineCell{Code}{isoDim} & \text{dimensionality of the heavy field under } $SU(2)_L$ \\
		\rowcolor[gray]{.92}
		\mmaInlineCell{Code}{hyperCharge} & \text{hyper-charge of the heavy field under }$U(1)_Y$. \\
		\rowcolor[gray]{.92}
		\mmaInlineCell{Code}{spin} & \text{spin quantum number of the heavy field.} \\
		\rowcolor[gray]{.92}
		\mmaInlineCell{Code}{mass} & \text{variable representing the mass of the heavy field} \\
		\hline
	\end{tabular}
\end{table}


Now that our field definitions are ready, it's time to write the Lagrangian in a form that the code understands. For that, first we have to create the representations of these fields in their component form. This is done by a specific function:
\begin{mmaCell}{Code}
		hfvecs = \mmaDef{defineHeavyFields}[\mmaDef{fields}]
\end{mmaCell}
{\scriptsize
\begin{mmaCell}{Output}
		\{\{\{ph[1,1], ph[1,2], ph[1,3]\}\}\}
\end{mmaCell}
}
To write the Lagrangian in a compact form one can define the heavy field as:
\begin{mmaCell}{Input}
		\mmaUnd{\(\pmb{\Phi}\)} = \mmaDef{hfvecs}[[1,1]]
\end{mmaCell}
{\scriptsize
\begin{mmaCell}{Output}
		\{ph[1,1], ph[1,2], ph[1,3]\}
\end{mmaCell}
}
With this definition,  the working Lagarangian,  given in regular form in Eq.~(\ref{lag1}), can be written as:\\
either in compact form like:
\begin{mmaCell}[morefunctionlocal={a}]{Input}
		LBSM = 2 \mmaUnd{\(\pmb{\kappa}\)} *Table[(\mmaDef{dag}[H].\mmaDef{tau}[a].\mmaDef{H}),\{a,3\}].\mmaUnd{\(\pmb{\Phi}\)} 
					- \mmaUnd{\(\pmb{\eta}\)} \mmaSup{\mmaDef{abs}[\mmaDef{H}]}{2}*(\mmaUnd{\(\pmb{\Phi}\)}.\mmaUnd{\(\pmb{\Phi}\)}) - \mmaFrac{1}{4}\mmaUnd{\(\pmb{\lambda}\)} \mmaSup{(\mmaUnd{\(\pmb{\Phi}\)}.\mmaUnd{\(\pmb{\Phi}\)})}{2};
\end{mmaCell}
%

\subsubsection{Other examples of Defining Fields}\label{sec:buildheavy}

So far we have mentioned how to integrate out a single heavy field. In case of multiple fields, the required format of the `\texttt{fieldList}' would look like
\begin{mmaCell}[label={}]{Code}
		fieldList =
		{
			{ \mmaPat{fieldName1},\mmaPat{components1},\mmaPat{colorDim1},\mmaPat{isoDim1},\mmaPat{hyperCharge1},\mmaPat{spin1},\mmaPat{mass1} },
			{ \mmaPat{fieldName2},\mmaPat{components2},\mmaPat{colorDim2},\mmaPat{isoDim2},\mmaPat{hyperCharge2},\mmaPat{spin2},\mmaPat{mass2} },
			...
			...
		};
\end{mmaCell}
Now we can create the heavy fields' list as:
\begin{mmaCell}[label={In[..]:=}]{Input}
		hfVecs = \mmaDef{defineHeavyFields}[fieldList]
\end{mmaCell}

In general, if we have $n$ number of heavy fields, then \mmaInlineCell{Code}{hfVecs} is a list, whose $i^{th}$ element is the representation of the $i^{th}$ heavy field.
Let us describe different cases in detail, where we have fields of different characteristics. A case with a single field is already shown in sec. \ref{sec:buildEx}.

Following the same proposal, we can define multiple heavy fields. A possible example is:
\begin{mmaCell}[index=1]{Code}
		fieldList2 =
		{
			{hf1, 3, 1, 3, 0, 0, m},
			{hf2, 6, 3, 2, 1/6, 0, m}
		};
\end{mmaCell}
with which we define
\begin{mmaCell}{Code}
		hfarray2 = \mmaDef{defineHeavyFields}[\mmaDef{fieldList2}]
\end{mmaCell}
{\scriptsize
\begin{mmaCell}{Output}
		\{\{\{hf1[1,1], hf1[1,2], hf1[1,3]\}\},
		  \{\{hf2[1,1] + i ihf2[1,1], hf2[1,2] + i ihf2[1,2]\},
		    \{hf2[2,1] + i ihf2[2,1], hf2[2,2] + i ihf2[2,2]\},
		    \{hf2[3,1] + i ihf2[3,1], hf2[3,2] + i ihf2[3,2]\}\}\}
\end{mmaCell}
}
Here, the representation for the first heavy field is:
\begin{mmaCell}{Input}
		hfvec1 = \mmaDef{hfarray2}[[1,1]]
\end{mmaCell}
{\scriptsize
\begin{mmaCell}{Output}
		\{hf1[1,1],hf1[1,2],hf1[1,3]\}
\end{mmaCell}
}
and the second field is represented as:
\begin{mmaCell}{Input}
		hfvec2 = \mmaDef{hfarray2}[[2]]
\end{mmaCell}
{\scriptsize
\begin{mmaCell}{Output}
		\{\{hf2[1,1] + i ihf2[1,1],hf2[1,2]+i ihf2[1,2]\},
		  \{hf2[2,1] + i ihf2[2,1],hf2[2,2]+i ihf2[2,2]\},
		  \{hf2[3,1] + i ihf2[3,1],hf2[3,2]+i ihf2[3,2]\}\}
\end{mmaCell}
}
Now, these two field representations (\mmaInlineCell{Code}{hfvec1} and \mmaInlineCell{Code}{hfvec2}) can be used to build the required Lagrangian.

For a spin-1 field, the field definition will be:
\begin{mmaCell}[index=1]{Input}
		fieldList3=
		\{
		  \{hf3,3,1,3,0,1,m3\}
		\};
\end{mmaCell}
We do not count the Lorentz components of a heavy field while writing the total number of field components (the second entry in \mmaInlineCell{Code}{fieldList}).
\begin{mmaCell}{Input}
		hfarray3= \mmaDef{defineHeavyFields}[\mmaDef{fieldList3}]
\end{mmaCell}
{\scriptsize
\begin{mmaCell}{Output}
	\{\{\{hf3[1][1],hf3[1][2],hf3[1][3],hf3[1][4]\},
	  \{hf3[2][1],hf3[2][2],hf3[2][3],hf3[2][4]\},
	  \{hf3[3][1],hf3[3][2],hf3[3][3],hf3[3][4]\}\}\}
\end{mmaCell}
}
\begin{mmaCell}{Input}
		hfvec3 = \mmaDef{hfarray3}[[1]]
\end{mmaCell}
{\scriptsize
\begin{mmaCell}{Output}
		\{\{hf3[1][1],hf3[1][2],hf3[1][3],hf3[1][4]\},
		  \{hf3[2][1],hf3[2][2],hf3[2][3],hf3[2][4]\},
		  \{hf3[3][1],hf3[3][2],hf3[3][3],hf3[3][4]\}\}
\end{mmaCell}
}

In a similar manner, the spin-1/2 field is represented as:
\begin{mmaCell}[index=1]{Input}
		fieldList4 =
		\{
		  \{hf4,3,1,3,0,\mmaFrac{1}{2},m4\}
		\};
\end{mmaCell}
As before, we do not count the Lorentz components of a heavy field.
\begin{mmaCell}{Input}
		hfarray4= \mmaDef{defineHeavyFields}[\mmaDef{fieldList4}]
\end{mmaCell}
{\scriptsize
\begin{mmaCell}{Output}
		\{\{\{\{\{hf4[1,1][1],hf4[1,1][2],hf4[1,1][3],hf4[1,1][4]\},
		        \{hf4b[1,1][1],hf4b[1,1][2],hf4b[1,1][3],hf4b[1,1][4]\}\},
		      \{\{hf4[1,2][1],hf4[1,2][2],hf4[1,2][3],hf4[1,2][4]\},
		        \{hf4b[1,2][1],hf4b[1,2][2],hf4b[1,2][3],hf4b[1,2][4]\}\},
		      \{\{hf4[1,3][1],hf4[1,3][2],hf4[1,3][3],hf4[1,3][4]\},
		        \{hf4b[1,3][1],hf4b[1,3][2],hf4b[1,3][3],hf4b[1,3][4]\}\}\}\}\}
\end{mmaCell}
}
\begin{mmaCell}{Input}
		hfvec4 = \mmaDef{hfarray4}[[1,1]][[All,1]]
\end{mmaCell}
{\scriptsize
\begin{mmaCell}{Output}
		\{\{hf4[1,1][1],hf4[1,1][2],hf4[1,1][3],hf4[1,1][4]\},
		  \{hf4[1,2][1],hf4[1,2][2],hf4[1,2][3],hf4[1,2][4]\},
		  \{hf4[1,3][1],hf4[1,3][2],hf4[1,3][3],hf4[1,3][4]\}\}
\end{mmaCell}
}
\begin{mmaCell}{Input}
		hfvec4b = \mmaDef{hfarray4}[[1,1]][[All,2]]
\end{mmaCell}
{\scriptsize
\begin{mmaCell}{Output}
		\{\{hf4b[1,1][1],hf4b[1,1][2],hf4b[1,1][3],hf4b[1,1][4]\},
		  \{hf4b[1,2][1],hf4b[1,2][2],hf4b[1,2][3],hf4b[1,2][4]\},
		  \{hf4b[1,3][1],hf4b[1,3][2],hf4b[1,3][3],hf4b[1,3][4]\}\}
\end{mmaCell}
}
Use \mmaInlineCell{Code}{hfvec4} and \mmaInlineCell{Code}{hfvec4b} as the field representation of the fermion (say, $\psi$) and its Lorentz conjugate ($\overline{\psi}$).

\subsection{How to run the code}\label{sec:howtorun}

\begin{table}[ht]
	\caption{Available functions for calculating Wilson Coefficients (WCs) for a given BSM Lagrangian}\label{tab:mainFuncs}
	\begin{center}
		\renewcommand*{\arraystretch}{2}
		\begin{tabular}{|cl|}
			\rowcolor[gray]{.92}
			\hline
			Function &  Details \\ 
			\hline
			\rowcolor[gray]{.92}
			\mmaInlineCell{Code}{treeOutput} & \text{Calculates tree level Wilson Coefficients.} \\
			\rowcolor[gray]{.92}
			\mmaInlineCell{Code}{loopOutput} & \text{Calculates 1 loop level Wilson Coefficients.} \\
			\rowcolor[gray]{.92}
			\mmaInlineCell{Code}{codexOutput} & \text{Generic function for WCs calculation upto 1 loop.} \\
			\hline
		\end{tabular}
	\end{center}
\end{table}

We have demonstrated how to build the Lagrangian in the previous section. Here we will discuss the necessary steps that needs to be followed to compute the Wilson coefficients (WC). Of the various options that \mmaInlineCell{Input}{\mmaDef{CoDEx}} provides to the users, the first is choice of operator bases (\mmaInlineCell{Code}{operBasis}) between (i) \mmaInlineCell{Code}{"Warsaw"}, and (ii) \mmaInlineCell{Code}{"SILH"}. Here \mmaInlineCell{Code}{"Warsaw"} is the default one. The next option is the level at which the user wants the WCs.\\

One can compute upto 1-loop WCs using this code.

\subsubsection{Tree Level}

To obtain the tree level Wilson coefficients, one needs to use the function \mmaInlineCell{Code}{treeOutput}, used as:
\begin{mmaCell}[label={}]{Code}
		\mmaDef{treeOutput}[\mmaPat{lagrangian},\mmaPat{fieldList}]
\end{mmaCell}
This will generate the WCs in the \mmaInlineCell{Code}{"Warsaw"} basis. Now, to compute the same in \mmaInlineCell{Code}{"SILH"} basis, one has to simply provide explicit choice of the operator basis as:
\begin{mmaCell}[label={}]{Code}
		\mmaDef{treeOutput}[\mmaPat{lagrangian},\mmaPat{fieldList},operBasis->"SILH"]
\end{mmaCell}

\subsubsection{1-loop Level}\label{sec:1loop}

To compute the WCs at 1-loop level only, we have to use another function, \mmaInlineCell{Code}{loopOutput}. This can be used as:
\begin{mmaCell}[label={}]{Code}
		\mmaDef{loopOutput}["model",\mmaPat{lagrangian},\mmaPat{fieldList}]
\end{mmaCell}
Unlike the tree level case, here we need the transformation property of the heavy fields under the given gauge symmetry. More precisely, the structure of the generators determined by the dimensionality of the heavy field's representations must be provided explicitly. We have provided their structure up to fundamental and quadruplet for $SU(3)_C$ and $SU(2)_L$ gauge groups respectively.
For more exotic BSM particles, we have kept provision for the user to define it. To do so, one has to run a function as:
\begin{mmaCell}[label={}]{Code}
		\mmaDef{initializeLoop}["model",\mmaPat{fieldList}]
\end{mmaCell}
If the dimensionality of the heavy fields are within the mentioned ranges then one does not need to provide the explicit structures of the generators. Otherwise, she has to provide all the generators explicitly for each and every heavy field as: 
\mmaInlineCell{Code}{isomodel[p,i]} and \mmaInlineCell{Code}{colmodel[p, a]} where `\mmaInlineCell{Code}{p}' denotes the number of heavy fields, and `\mmaInlineCell{Code}{i}' and `\mmaInlineCell{Code}{a}' run from 1 to 3 and 1 to 8 respectively.

\subsubsection{One Function to find them All}

\begin{table}[ht]
	\caption{The Options for \texttt{codexOutput}. Other than these, this function also takes all Options of \texttt{formPick}}\label{tab:codexOutput}
			\begin{center}
	\renewcommand*{\arraystretch}{2}
	\begin{tabular}{|ccl|}
		\rowcolor[gray]{.92}
		\hline
		Option & Default Value & Details \\ 
		\hline\rowcolor[gray]{.92}
		\mmaInlineCell{Code}{monitor} &  \mmaInlineCell{Code}{True} & \text{Shows an animation while computing} \\
		\rowcolor[gray]{.92}
		\mmaInlineCell{Code}{appearance} & \mmaInlineCell{Code}{"Percolate"} & \text{Appearence of the animation.} \\
		\rowcolor[gray]{.92}
		 & (\text{for} \text{Version}$\geq$ 11.) & \\
		\rowcolor[gray]{.92}
		\mmaInlineCell{Code}{operBasis} & \mmaInlineCell{Code}{"Warsaw"} & \text{Choice of basis of the  Dim.-6 operators } \\
		\rowcolor[gray]{.92}
		\mmaInlineCell{Code}{format} & \mmaInlineCell{Code}{List} & \text{The output format.} \\
		\rowcolor[gray]{.92}
		\mmaInlineCell{Code}{model} & \mmaInlineCell{Code}{""} & Takes the same input as \mmaInlineCell{Code}{initializeLoop}.\\
		\rowcolor[gray]{.92}
		 & & If left blank, will give the tree-level. \\
		\rowcolor[gray]{.92}
		\mmaInlineCell{Code}{outRange} & \mmaInlineCell{Code}{All} & The level at which output is evaluated.\\
		\rowcolor[gray]{.92}
		& &  `\mmaInlineCell{Code}{All}' means the result calculates both\\
		\rowcolor[gray]{.92}
		& &  tree and loop level results and combines them. \\
		\rowcolor[gray]{.92}
		\mmaInlineCell{Code}{ibp} &  \mmaInlineCell{Code}{True} & Turns the `Integration by Parts' option on.\\
		\rowcolor[gray]{.92}
		& &  Detail available on the \mmaInlineCell{Code}{loopOutput} page.\\
		  \hline
	\end{tabular}
		\end{center}
\end{table}

Now, one can wish to get all the WCs, i.e., tree and 1-loop levels together. For that one can simply use the following function as: 
\begin{mmaCell}[label={}]{Input}
		codexOutput[\mmaPat{lagrangian},\mmaPat{fieldList},model\(\pmb{\to}\)\mmaPat{modelName_String}]
\end{mmaCell}
Essentially, this function is all one needs to calculate the WCs and even format them in the correct way. With careful choice of \mmaInlineCell{Code}{OptionValues}, one can obtain results at different levels, with different operator bases, and in different formats. We have enlisted the main options which can be found using \mmaInlineCell{Code}{Options[codexOutput]}.
%

\subsection{RGEs of WCs -- Anomalous dimension matrix and choice of basis}\label{sec:rgflow}

For a given BSM Lagrangian we can compute the effective dimension-6 operators and the associated Wilson coefficients by integrating out the heavy fields. \mmaInlineCell{Input}{\mmaDef{CoDEx}} can provide these results in two different bases: \mmaInlineCell{Code}{"Warsaw"} and \mmaInlineCell{Code}{"SILH"}. In the \mmaInlineCell{Code}{"Warsaw"} basis, operators form a complete basis unlike the \mmaInlineCell{Code}{"SILH"} one. Thus we prefer to perform the running of the Wilson Coefficients in \mmaInlineCell{Code}{"Warsaw"} basis only. 
Once the WCs are computed at the high scale, one can run those effective operators using the Anomalous dimension matrix and can compute the operator structures at the electro-weak (EW) scale by using the \mmaInlineCell{Code}{RGFlow} function. Results of this module will help to connect the EW observables and the BSM physics through the effective  operators and the WCs. 

To perform the RG evolution of the WCs, we need to use the function : 
\begin{mmaCell}[label={}]{Input}
		RGFlow[\mmaPat{WCList_List},\mmaPat{MatchingScale},\mmaPat{\(\pmb{\mu}\)}]
\end{mmaCell}

This function works in the following way:
\begin{itemize}
	\item \mmaInlineCell{Code}{RGFlow} takes a list as its first argument. List output from \mmaInlineCell{Code}{treeOuput}, \mmaInlineCell{Code}{loopOutput} and \mmaInlineCell{Code}{codexOutput} functions should be used as the first argument.
	\item \mmaInlineCell{Code}{RGFlow} takes Wilson Coefficients of dimension six operators in \mmaInlineCell{Code}{"Warsaw"} basis only.
	\item The second argument is the scale at which the full BSM theory is matched with the EFT.
	\item \mmaInlineCell{Input}{\mmaPat{\(\pmb{\mu}\)}} can be any energy scale below the matching scale.
\end{itemize}

Here, both the \mmaInlineCell{Code}{\mmaPat{MatchingScale}} and \mmaInlineCell{Input}{\mmaPat{\(\pmb{\mu}\)}} can be symbolic inputs. The working principle of this function works is as follows:
\begin{itemize}
	\item Load the package through:
\begin{mmaCell}[index=1]{Code}
		Needs["CoDEx`"]
\end{mmaCell}
	\item Say that the following is the output in the form of list of WCs, that you had found from an earlier session of \mmaInlineCell{Input}{\mmaDef{CoDEx}} and had saved. Let's give it a name:
\begin{mmaCell}{Input}
		trrt=
		\{
			\{"qH",-\mmaFrac{\mmaUnd{\(\pmb{\eta}\)} \mmaSup{\mmaUnd{\(\pmb{\kappa}\)}}{2}}{\mmaSup{m}{4}}\}, \{"qHbox",\mmaFrac{\mmaSup{\mmaUnd{\(\pmb{\kappa}\)}}{2}}{\mmaSup{m}{4}}\},\{"qHD",-\mmaFrac{2 \mmaSup{\mmaUnd{\(\pmb{\kappa}\)}}{2}}{\mmaSup{m}{4}}\}
		\};
\end{mmaCell}
	These WCs are evaluated at the high scale. Now, to compute the WCs at the electro-weak scale(\mmaInlineCell{Input}{\mmaPat{mu}}) we need to perform their RGEs.
	After setting the matching scale (high scale) at the mass of the heavy particle (`\mmaInlineCell{Code}{\mmaUnd{m}}'), we have to recall the function \mmaInlineCell{Code}{RGFlow} as:
\begin{mmaCell}{Input}
		floRes1 = \mmaDef{RGFlow}[\mmaDef{trrt},m,\mmaUnd{\(\pmb{\mu}\)}]
\end{mmaCell}
{\scriptsize
\begin{mmaCell}{Output}
		\{
			\{qH,-\mmaFrac{\(\eta\) \mmaSup{\(\kappa\)}{2}}{\mmaSup{m}{4}} + \mmaFrac{3 \mmaSup{gW}{2} \mmaSup{\(\kappa\)}{2} Log[\mmaFrac{\(\mu\)}{m}]}{4	\mmaSup{m}{4} \mmaSup{\(\pi\)}{2}} + \mmaFrac{3 \mmaSup{gW}{4} \mmaSup{\(\kappa\)}{2} Log[\mmaFrac{\(\mu\)}{m}]}{32 \mmaSup{m}{4} \mmaSup{\(\pi\)}{2}} - \mmaFrac{3 \mmaSup{gY}{2} \mmaSup{\(\kappa\)}{2} Log[\mmaFrac{\(\mu\)}{m}]}{4 \mmaSup{m}{4} \mmaSup{\(\pi\)}{2}} + 
				\mmaFrac{3 \mmaSup{gW}{2} \mmaSup{gY}{2} \mmaSup{\(\kappa\)}{2} Log[\mmaFrac{\(\mu\)}{m}]}{16 \mmaSup{m}{4} \mmaSup{\(\pi\)}{2}}+\mmaFrac{3 \mmaSup{gY}{4} \mmaSup{\(\kappa\)}{2} Log[\mmaFrac{\(\mu\)}{m}]}{32 \mmaSup{m}{4} \mmaSup{\(\pi\)}{2}} + \mmaFrac{27 \mmaSup{gW}{2} \(\eta\) \mmaSup{\(\kappa\)}{2} Log[\mmaFrac{\(\mu\)}{m}]}{32 \mmaSup{m}{4} \mmaSup{\(\pi\)}{2}} +
				\mmaFrac{9 \mmaSup{gY}{2} \(\eta\) \mmaSup{\(\kappa\)}{2} Log[\mmaFrac{\(\mu\)}{m}]}{32 \mmaSup{m}{4} \mmaSup{\(\pi\)}{2}}+\mmaFrac{5 \mmaSup{gW}{2} \mmaSup{\(\kappa\)}{2}	\(\lambda\) Log[\mmaFrac{\(\mu\)}{m}]}{6 \mmaSup{m}{4} \mmaSup{\(\pi\)}{2}}\},
			\{qHbox,\mmaFrac{\mmaSup{\(\kappa\)}{2}}{\mmaSup{m}{4}}-\mmaFrac{\mmaSup{gW}{2} \mmaSup{\(\kappa\)}{2} Log[\mmaFrac{\(\mu\)}{m}]}{4 \mmaSup{m}{4} \mmaSup{\(\pi\)}{2}}-\mmaFrac{7 \mmaSup{gY}{2} \mmaSup{\(\kappa\)}{2} Log[\mmaFrac{\(\mu\)}{m}]}{24	\mmaSup{m}{4} \mmaSup{\(\pi\)}{2}}\},
			\{qHD,-\mmaFrac{2 \mmaSup{\(\kappa\)}{2}}{\mmaSup{m}{4}}-\mmaFrac{9 \mmaSup{gW}{2} \mmaSup{\(\kappa\)}{2} Log[\mmaFrac{\(\mu\)}{m}]}{16 \mmaSup{m}{4} \mmaSup{\(\pi\)}{2}} + \mmaFrac{25 \mmaSup{gY}{2} \mmaSup{\(\kappa\)}{2} Log[\mmaFrac{\(\mu\)}{m}]}{48 \mmaSup{m}{4} \mmaSup{\(\pi\)}{2}}\},
			\{qeH[1,1],\mmaFrac{19 \mmaSup{gW}{2} \mmaSup{\(\kappa\)}{2} Log[\mmaFrac{\(\mu\)}{m}] \mmaSup{Yu}{\(\dagger\)}[e]}{48 \mmaSup{m}{4} \mmaSup{\(\pi\)}{2}}-\mmaFrac{3 \mmaSup{gY}{2} \mmaSup{\(\kappa\)}{2} Log[\mmaFrac{\(\mu\)}{m}] \mmaSup{Yu}{\(\dagger\)}[e]}{16 \mmaSup{m}{4} \mmaSup{\(\pi\)}{2}}\},
			\{quH[1,1],\mmaFrac{19 \mmaSup{gW}{2} \mmaSup{\(\kappa\)}{2} Log[\mmaFrac{\(\mu\)}{m}] \mmaSup{Yu}{\(\dagger\)}[u]}{48 \mmaSup{m}{4} \mmaSup{\(\pi\)}{2}}-\mmaFrac{3 \mmaSup{gY}{2} \mmaSup{\(\kappa\)}{2} Log[\mmaFrac{\(\mu\)}{m}] \mmaSup{Yu}{\(\dagger\)}[u]}{16 \mmaSup{m}{4} \mmaSup{\(\pi\)}{2}}\},
			\{qdH[1,1],\mmaFrac{19 \mmaSup{gW}{2} \mmaSup{\(\kappa\)}{2} Log[\mmaFrac{\(\mu\)}{m}] \mmaSup{Yu}{\(\dagger\)}[d]}{48 \mmaSup{m}{4} \mmaSup{\(\pi\)}{2}}-\mmaFrac{3 \mmaSup{gY}{2} \mmaSup{\(\kappa\)}{2} Log[\mmaFrac{\(\mu\)}{m}] \mmaSup{Yu}{\(\dagger\)}[d]}{16 \mmaSup{m}{4} \mmaSup{\(\pi\)}{2}}\},
			\{q1Hl[1,1],\mmaFrac{\mmaSup{gY}{2} \mmaSup{\(\kappa\)}{2} Log[\mmaFrac{\(\mu\)}{m}]}{96 \mmaSup{m}{4} \mmaSup{\(\pi\)}{2}}\},	\{q3Hl[1,1],\mmaFrac{\mmaSup{gW}{2} \mmaSup{\(\kappa\)}{2} Log[\mmaFrac{\(\mu\)}{m}]}{96 \mmaSup{m}{4} \mmaSup{\(\pi\)}{2}}\},
			\{qHe[1,1],\mmaFrac{\mmaSup{gY}{2} \mmaSup{\(\kappa\)}{2} Log[\mmaFrac{\(\mu\)}{m}]}{48 \mmaSup{m}{4} \mmaSup{\(\pi\)}{2}}\}, \{q1Hq[1,1],-\mmaFrac{\mmaSup{gY}{2} \mmaSup{\(\kappa\)}{2} Log[\mmaFrac{\(\mu\)}{m}]}{288 \mmaSup{m}{4} \mmaSup{\(\pi\)}{2}}\},
			\{q3Hq[1,1],\mmaFrac{\mmaSup{gW}{2} \mmaSup{\(\kappa\)}{2} Log[\mmaFrac{\(\mu\)}{m}]}{96 \mmaSup{m}{4} \mmaSup{\(\pi\)}{2}}\}, \{qHu[1,1],-\mmaFrac{\mmaSup{gY}{2} \mmaSup{\(\kappa\)}{2} Log[\mmaFrac{\(\mu\)}{m}]}{72 \mmaSup{m}{4} \mmaSup{\(\pi\)}{2}}\},
			\{qHd[1,1],\mmaFrac{\mmaSup{gY}{2} \mmaSup{\(\kappa\)}{2} Log[\mmaFrac{\(\mu\)}{m}]}{144 \mmaSup{m}{4} \mmaSup{\(\pi\)}{2}}\}
		\}
\end{mmaCell}
}	
	\item One can reformat, save, and/or export all these WCs corresponding to the effective operators at the electro-weak scale (\mmaInlineCell{Input}{\mmaUnd{\(\pmb{\mu}\)}}) to \LaTeX, using \mmaInlineCell{Input}{formPick}. Below is an illustrative example:
\begin{mmaCell}{Input}
		\mmaDef{formPick}["Warsaw","Detailed2",\mmaDef{floRes1},Frame\(\pmb{\to}\)All,FontSize\(\pmb{\to}\)\mmaDef{Medium},
		FontFamily\(\pmb{\to}\)"Times New Roman"]
\end{mmaCell}
\begin{mmaCell}{Output}
	    
\end{mmaCell}
\begin{center}
	{\tiny \begin{tabular}{|*{3}{c|}}
			\hline
			$Q_H$  &  $\left(H^{\dagger }H^{  })^3\right.$  &  $\frac{\log \left(\frac{\mu }{m}\right) \left(-\frac{\eta  \kappa ^2 \left(-\frac{27 \text{gW}^2}{2}-\frac{9 \text{gY}^2}{2}\right)}{m^4}- \ldots +\frac{40 \text{gW}^2 \kappa ^2 \lambda }{3 m^4}\right)}{16 \pi ^2}-\frac{\eta  \kappa ^2}{m^4}$  \\
			\hline
			$\vdots$ & $\vdots$ & $\vdots$ \\
			\hline
			$Q_{\text{Hd}}$  &  $\left(H^{\dagger }\text{ i }\overleftrightarrow{\mathcal{D}  }_{\mu } H^{  }\text{)(}\bar{d} \gamma ^{\mu }\text{ d)}\right.$  &  $\frac{\text{gY}^2 \kappa ^2 \log \left(\frac{\mu }{m}\right)}{144 \pi ^2 m^4}$\\
			\hline
		\end{tabular}
	}
\end{center}
	Here we have shown a truncated version of the resulting long table\footnote{Fun-fact: This table in \LaTeX~format, and other similar results used in this draft, are all created using \texttt{formPick} as well.}.
	\item Remember that the RGE of WCs can only be performed in the \mmaInlineCell{Input}{"Warsaw"} basis (as it a complete one) and not in the \mmaInlineCell{Input}{"SILH"} basis.
\begin{mmaCell}{Input}
		RGFlow[\{\{"oH",\mmaFrac{\mmaSup{\mmaUnd{\(\pmb{\kappa}\)}}{2}}{\mmaSup{m}{4}}\}\},m,\mmaUnd{\(\pmb{\mu}\)}]
\end{mmaCell}
{\scriptsize
\begin{mmaCell}{Output}
		RG flow only works when the Wilson Coefficient basis is complete. For now, it only works for the `Warsaw' basis.
\end{mmaCell}
}
\end{itemize}

\subsection{Detailed example: Electro-weak $SU(2)_L$ Real Singlet Scalar}\label{sec:detailex}
Here we demonstrate the work-flow of \mmaInlineCell{Code}{\mmaDef{CoDEx}} with the help of a complete analysis of a representative model. This and many others are listed in the package documentation. We also list the results of the other models in the appendix of this draft \ref{sec:app1}.
Say the Lagrangian is:
{\small
\begin{align}
\label{lbsmRSS}
\mathcal{L}_{BSM} & =   \mathcal{L}_{SM} \ + \ \frac{1}{2} \ (\mathcal{\partial}_{\mu} \phi )^{2} \ - \ \frac{1}{2} \ m_{\phi}^{2} \ \phi^{2} - c_{a} |H|^{2} \phi  - \frac{1}{2} \kappa |H|^{2}  \phi^{2} - \frac{1}{3!} \ \mu \phi^{3}  - \frac{1}{4!} \ \lambda_{\phi} \phi^{4}\,.
\end{align}
}
Here $\phi$ is the real singlet scalar. Once this field is integrated out, few effective operators will emerge. To obtain those effective dimension-6 operators and their respective  Wilson Coefficients using \mmaInlineCell{Code}{\mmaDef{CoDEx}}, we need to perform the following steps:

\begin{table*}[ht]
	\caption{Effective operators and Wilson Coefficients for Real Singlet Scalar.}
	\label{tab:realSing}
	\small
	\centering
	\renewcommand{\arraystretch}{2.5}
	\subfloat[``SILH'' basis]{
	\begin{tabular}{|c|c|}
	\hline
	$O_6$  &  $\frac{c_{a}^3 \mu }{6 m_{\phi}^6}-\frac{c_{a}^2 \kappa  \lambda_{\phi} }{32 \pi ^2 m_{\phi}^4}-\frac{c_{a}^2 \kappa }{2 m_{\phi}^4}-\frac{\kappa ^3}{192 \pi ^2 m_{\phi}^2}$  \\
	\hline
	$O_H$  &  $\frac{c_{a}^2 \mu ^2}{192 \pi ^2 m_{\phi}^6}+\frac{c_{a}^2 \lambda_{\phi} }{16 \pi ^2 m_{\phi}^4}+\frac{c_{a}^2}{m_{\phi}^4}-\frac{c_{a} \kappa  \mu }{96 \pi ^2 m_{\phi}^4}+\frac{\kappa ^2}{192 \pi ^2 m_{\phi}^2}$  \\
	\hline
\end{tabular}\label{tab:realSingS}		
}
\subfloat[``Warsaw'' basis]{
	\begin{tabular}{|c|c|}
	\hline
	$Q_H$  &  $\frac{c_{a}^3 \mu }{6 m_{\phi}^6}-\frac{c_{a}^2 \kappa  \lambda_{\phi} }{32 \pi ^2 m_{\phi}^4}-\frac{c_{a}^2 \kappa }{2 m_{\phi}^4}-\frac{\kappa ^3}{192 \pi ^2 m_{\phi}^2}$  \\
	\hline
	$Q_{\text{HD}}$  &  $\frac{c_{a}^2 \mu ^2}{96 \pi ^2 m_{\phi}^6}-\frac{c_{a}^2 \lambda_{\phi} }{8 \pi ^2 m_{\phi}^4}-\frac{2 c_{a}^2}{m_{\phi}^4}-\frac{c_{a} \kappa  \mu }{48 \pi ^2 m_{\phi}^4}+\frac{\kappa ^2}{96 \pi ^2 m_{\phi}^2}$  \\
	\hline
\end{tabular}\label{tab:realSingW}
}
\end{table*}

\begin{enumerate}
	\item First, load the package:
\begin{mmaCell}[index=1]{Input}
		Needs["CoDEx`"]
\end{mmaCell}
	\item We have to define the field $\phi$ as:
\begin{mmaCell}[index=1]{Input}
		fieldewrss=
		\{
		  \{hf,1,1,1,0,0,m\}
		\};
\end{mmaCell}
\begin{mmaCell}{Input}
		hfvecsewrss=\mmaDef{defineHeavyFields}[\mmaDef{fieldewrss]};
\end{mmaCell}
\begin{mmaCell}{Input}
		\mmaUnd{\(\pmb{\phi}\)} = \mmaDef{hfvecsewrss}[[1,1,1]]
\end{mmaCell}
\begin{mmaCell}{Output}
		hf[1,1]
\end{mmaCell}

	\item Then we need to build the relevant part of the  Lagrangian (involving the heavy field only). As a side-note, we should mention, that for \mmaInlineCell{Input}{\mmaDef{CoDEx}} to function, it does not need the heavy field kinetic term (the covariant derivative and the mass terms). Thus, the only part of the Lagrangian we need here is:
\begin{mmaCell}{Input}
		Lpotenewrss=Expand[-\mmaSub{c}{a}*\mmaSup{\mmaDef{abs}[\mmaDef{H}]}{2}*\mmaDef{\(\pmb{\phi}\)}-\mmaFrac{\mmaUnd{\(\pmb{\kappa}\)}}{2}*\mmaSup{\mmaDef{abs}[\mmaDef{H}]}{2}*\mmaSup{\mmaDef{\(\pmb{\phi}\)}}{2}-\mmaFrac{1}{3!} \mmaUnd{\(\pmb{\mu}\)}*\mmaSup{\mmaDef{\(\pmb{\phi}\)}}{3}-\mmaFrac{1}{4!} \mmaUnd{\(\pmb{\lambda}\)}*\mmaSup{\mmaDef{\(\pmb{\phi}\)}}{4}];
\end{mmaCell}
	\item Next, we need to construct the symmetry generators:
\begin{mmaCell}{Input}
		\mmaDef{initializeLoop}["ewrss",\mmaDef{fieldewrss},printInfo\(\pmb{\to}\)False]
\end{mmaCell}
{\small
\begin{mmaCell}[label={\mmaShd{>>}}]{Output}
		Isospin Symmetry Generators for the field `hf' are isoewrss[1,a] = 0
\end{mmaCell}
\begin{mmaCell}[label={\mmaShd{>>}}]{Output}
		Color Symmetry Generators for the field `hf' are colewrss[1,a] = 0
\end{mmaCell}
}
(See the documentation of \mmaInlineCell{Input}{initializeLoop} for details.)
	\item The last step is the computation of effective operators and associated WCs as:
\begin{mmaCell}{Input}
		res1=\mmaDef{codexOutput}[\mmaDef{Lpotenewrss},\mmaDef{fieldewrss},model\(\pmb{\to}\)"ewrss"];
		\mmaDef{formPick}["Warsaw","Detailed2",\mmaDef{res1},FontSize\(\pmb{\to}\)\mmaDef{Medium},
				FontFamily\(\pmb{\to}\)"Times New Roman",Frame\(\pmb{\to}\)All]
\end{mmaCell}
	\item The output is obtained in \mmaInlineCell{Code}{"Warsaw"} basis and is formatted as a detailed table in \mmaInlineCell{Code}{TraditionalForm}. There is provision to export the result in \LaTeX~ format. Table \ref{tab:realSingW} is actually obtained from the output of the code above.
We can compute the same in \mmaInlineCell{Code}{"SILH"} basis as well and for that we have to use:
\begin{mmaCell}{Input}
		res2=\mmaDef{codexOutput}[\mmaDef{Lpotenewrss},\mmaDef{fieldewrss},model\(\pmb{\to}\)"ewrss",operBasis\(\pmb{\to}\)"SILH"];
		\mmaDef{formPick}["SILH","Detailed2",\mmaDef{res2},FontSize\(\pmb{\to}\)\mmaDef{Medium},
				FontFamily\(\pmb{\to}\)"Times New Roman",Frame\(\pmb{\to}\)All]
\end{mmaCell}
	Output of this can be found in table \ref{tab:realSingS}.
	\item As is demonstrated in section \ref{sec:rgflow}, these resulting WCs can then be runned down to the electro-weak scale, using \mmaInlineCell{Code}{RGFlow}.
\end{enumerate}

\subsection{Miscellaneous}

\mmaInlineCell{Code}{\mmaDef{CoDEx}} is written in {\em Wolfram Language$^\text{\tiny \textregistered}$} \cite{Mathematica}. Careful steps has been taken to speed-up the code using parallelization over multi-cores, when available, while keeping the customizability for the user. All the example models listed in this draft and in the documentation have been run on different processors, with different operating systems and versions of Mathematica
.

When run on a 1.6 GHz Intel$^\text{\tiny \textregistered}$ Core i5 processor,
the models take $\sim$ 20 to 2000 seconds to run. Tree level runs never take more than a minute. The only exception is 2HDM, which we have run in a 16 core Xeon processor, with 32 GB RAM. 

How much time it takes to get the WCs for the user's model, depends on its structure and complexity. We hope a user can have a clear idea about run time if she runs all the examples given in documentation as trials.

\section{Summary}

\mmaInlineCell{Input}{\mmaDef{CoDEx}} allows one to integrate out single and(or) multiple degenerate heavy field(s) in a gauge covariant way. The user needs to provide the part of the Lagrangian that involves heavy BSM fields only.  She needs to identify that BSM field by providing its no. of component fields, spin, mass and quantum numbers under Standard Model gauge symmetry in a certain way. \mmaInlineCell{Code}{\mmaDef{CoDEx}} then integrates out the heavy field propagators from all tree level and(or) 1-loop processes, and generates the Wilson Coefficients for an exhaustive set of effective operators in both \mmaInlineCell{Code}{"SILH"} and \mmaInlineCell{Code}{"Warsaw"} bases.  It allows the user to run down the operators in \mmaInlineCell{Code}{"Warsaw"} basis to the electro-weak scale.  
As the precision observables can be recast in terms of the effective operators, it will be really helpful to test the BSM physics under the light of electro-weak precision data. 
A list of example models are provided along with the package. These include a variety of field representations usually used by the BSM model builders. 

In the present version of \mmaInlineCell{Code}{\mmaDef{CoDEx}}, the user can compute up to mass dimension-6 operators by integrating out up to spin-1 particles. 
This package can integrate out only heavy field propagators at the tree and 1-loop levels. In a future version, we will include a few other aspects, such that it can deal with  (i) loops containing light (SM) - heavy (BSM) mixed field propagators, (ii) non-degenerate multiple heavy BSM  fields \cite{our-future-paper}.

\section{Acknowledgements}

The authors acknowledge the useful discussions with Soumitra Nandi.  This work is supported by the Department of Science and
Technology, Government of India, under the Grant IFA12/PH/34 (INSPIRE Faculty Award);  the Science and Engineering Research Board, Government of India, under the agreement
SERB/PHY/2016348 (Early Career Research Award), and Initiation Research Grant, agreement no. IITK/PHY/2015077, by IIT Kanpur.

\section*{References}

\bibliography{code}

\appendix

\section{Field representations}\label{sec:app1}

For the rest of the draft, we build and use Lagrangians using the fields listed in Table~\ref{allbsmfields}. We have checked that for these given models, the \mmaInlineCell{Code}{\mmaDef{CoDEx}} generated results in \mmaInlineCell{Code}{"SILH"} basis are in well agreement with those given in Refs.~\cite{Henning:2014wua,Henning:2015alf,Henning:2016lyp}.

\begin{table}[ht]
	\caption{SM gauge quantum numbers of BSM fields.}
	\label{allbsmfields}
	\small
	\centering
	\renewcommand{\arraystretch}{1.5}
	\begin{tabular}{|c|c|c|c|c|c|c|}
		\hline
		BSM & No. of & $SU(3)_{C}$ &  $SU(2)_{L}$ & $U(1)_{Y}$ & & \\
		Field & field & quantum &  quantum  & charge\footnote{Electromagnetic charge $Q = T_{3} + Y$, where $T_{3}$ is isospin quantum number. }  & spin & mass \\
		& components & number &  number &  &  &  \\
		\hline
		$\phi$ & 1 & 1 & 1 & 0 & 0 & $m_{\phi}$ \\
		\hline
		$\Phi$ & 3 & 1 & 3 & 0 & 0 & $m_{\Phi}$ \\
		\hline
		$\Delta$ & 3 & 1 & 3 & 1 & 0 & $m_{\Delta}$ \\
		\hline
		$\Theta$ & 4 & 1 & 4 &  $3/2$  & 0 & $m_{\Theta}$ \\
		\hline
		$\psi$ & 1 & 1 & 1 &  0  & 1/2 & $m_{\psi}$ \\
		\hline
		$\Sigma$ & 3 & 1 & 3 &  0  & 1/2 & $m_{\Sigma}$ \\
		\hline
		Q & 3 & 1 & 3 &  0  & 1 & $m_{Q}$ \\
		\hline
		$\varphi$ & 2 & 1 & 2 &  - 1/2  & 0 & $m_{\varphi}$ \\
		\hline
		K & 1 & 1 & 1 &  0  & 1 & $m_{K}$ \\
		\hline
		$\tilde{Q}_{3L}$ & 6 & 3 & 2 &  1/6  & 0 & $m_{\tilde{Q}_{3}}$ \\
		\hline
		$\tilde{t}_{R}$ & 3 & 3 & 1 &  2/3  & 0 & $m_{\tilde{t}_{R}}$ \\
		\hline
	\end{tabular}
\end{table}

\section{Examples:  Single heavy BSM field}

\begin{enumerate}
	\item {\underline{Electro-weak $SU(2)_L$ Singlet Scalar with hypercharge $Y=0$}}: Discussed in detail in section \ref{sec:detailex}.
	\item {\underline{Electro-weak $SU(2)_L$ Triplet Scalar with hypercharge $Y=0$}}
{\small
	\begin{align}\label{lbsmRTS}
	\mathcal{L}_{BSM} = \ & \ \mathcal{L}_{SM} \ + \ \frac{1}{2} \ (\mathcal{D}_{\mu} \Phi )^{2} \ - \ \frac{1}{2} \ m_{\Phi}^{2} \ \Phi^{a} \ \Phi^{a} + 2 \ \kappa \ H^{\dagger} \tau^{a} H \ \Phi^{a} - \ \eta \ |H|^{2} \ \Phi^{a} \ \Phi^{a} - \frac{1}{4} \ \lambda_{\Phi} \ (\Phi^{a} \Phi^{a})^{2}.
	\end{align}
}
	Here, the heavy field is $\Phi$. The internal quantum numbers and its other required properties are given in Table~\ref{allbsmfields}. Once the $\Phi$ is integrated out using \mmaInlineCell{Input}{\mmaDef{CoDEx}}, the effective operators upto dimension-6 for both \mmaInlineCell{Code}{"SILH"} and \mmaInlineCell{Code}{"Warsaw"} bases are generated which are listed in Table~\ref{tab:RTS}.

\begin{table*}[ht]
	\caption{Effective operators and Wilson Coefficients in for Real Triplet Scalar (Y=0) model.}
	\label{tab:RTS}
	\small
	\centering
	\subfloat[``SILH'' basis]{
		\renewcommand{\arraystretch}{2}
		\begin{tabular}{|c|c|}
			\hline
			Dimension-6 operators & Wilson Coefficient \\
			\hline
			$O_{2W}$  &  $\frac{g_W^2}{480 \pi ^2 m_{\Phi}^2}$  \\
			\hline
			$O_{3W}$  &  $\frac{g_W^2}{480 \pi ^2 m_{\Phi}^2}$  \\
			\hline
			$O_6$  &  $-\frac{5 \eta  \kappa ^2 \lambda_{\Phi} }{8 \pi ^2 m_{\Phi}^4}-\frac{\eta  \kappa ^2}{m_{\Phi}^4}-\frac{\eta ^3}{8 \pi ^2 m_{\Phi}^2}$  \\
			\hline
			$O_H$  &  $\frac{\eta ^2}{16 \pi ^2 m_{\Phi}^2}$  \\
			\hline
			$O_R$  &  $\frac{5 \kappa ^2 \lambda_{\Phi} }{4 \pi ^2 m_{\Phi}^4}+\frac{2 \kappa ^2}{m_{\Phi}^4}$  \\
			\hline
			$O_T$  &  $\frac{5 \kappa ^2 \lambda_{\Phi} }{8 \pi ^2 m_{\Phi}^4}+\frac{\kappa ^2}{m_{\Phi}^4}$  \\
			\hline
			$O_{\text{WW}}$  &  $\frac{\eta }{96 \pi ^2 m_{\Phi}^2}$  \\
			\hline
		\end{tabular}
			}
	\subfloat[``Warsaw'' basis]{
		\renewcommand{\arraystretch}{2}
		\begin{tabular}{|c|c|}
			\hline
			Dimension-6 operators & Wilson Coefficient \\
			\hline
			$Q_H$  &  $-\frac{5 \eta  \kappa ^2 \lambda_{\Phi} }{8 \pi ^2 m_{\Phi}^4}-\frac{\eta  \kappa ^2}{m_{\Phi}^4}-\frac{\eta ^3}{8 \pi ^2 m_{\Phi}^2}$  \\
			\hline
			$Q_{H\square }$  &  $\frac{5 \kappa ^2 \lambda_{\Phi} }{8 \pi ^2 m_{\Phi}^4}+\frac{\kappa ^2}{m_{\Phi}^4}$  \\
			\hline
			$Q_{\text{HD}}$  &  $-\frac{5 \kappa ^2 \lambda_{\Phi} }{4 \pi ^2 m_{\Phi}^4}-\frac{2 \kappa ^2}{m_{\Phi}^4}+\frac{\eta ^2}{8 \pi ^2 m_{\Phi}^2}$  \\
			\hline
			$Q_{\text{HW}}$  &  $\frac{\eta  g_W^2}{96 \pi ^2 m_{\Phi}^2}$  \\
			\hline
			$Q_W$  &  $\frac{g_W^3}{2880 \pi ^2 m_{\Phi}^2}$  \\
			\hline
		\end{tabular}
			}
\end{table*}

	
	\item {\underline{Electro-weak $SU(2)_L$ Triplet Scalar with hypercharge $Y=1$}}
{\small 
	\begin{align}
	\label{lbsmCTS}
	\mathcal{L}_{BSM} = \ & \ \mathcal{L}_{SM} \ + Tr [ (\mathcal{D}_{\mu} \Delta)^{\dagger} (\mathcal{D}^{\mu} \Delta ) ] - m_{\Delta}^{2} Tr [ \Delta^{\dagger} \Delta ] + \mathcal{L}_{Y} - V(H,\Delta) 
	\end{align}
}	
	
	where,
{\small 
	\begin{align}
	V(H,\Delta) =& \zeta_{1} (H^{\dagger} H) Tr[ \Delta^{\dagger} \Delta ] + \zeta_{2} (H^{\dagger} \tau^{i} H) Tr[ \Delta^{\dagger} \tau^{i} \Delta ] + [ \mu ( H^{T} i \sigma^{2} \Delta^{\dagger} H ) + h.c.] \\
	\text{and}~~~~~\mathcal{L}_{Y} =& y_{\Delta} L^{T} C i \tau^2 \Delta L + h.c. 
	\end{align}
}
	Here, the heavy field is $\Delta$. The internal quantum numbers and its other required properties are given in Table~\ref{allbsmfields}. Once the $\Delta$ is integrated out using \mmaInlineCell{Input}{\mmaDef{CoDEx}}, the effective operators upto dimension-6 for both all bases are generated and are listed in Table~\ref{tab:CTS}.

\begin{table*}[!hbt]
	\caption{Effective operators and Wilson Coefficients for Complex Triplet Scalar (Y=1)  model.}
	\label{tab:CTS}
	\small
	\centering
	\renewcommand{\arraystretch}{2.5}
	\subfloat[``SILH'' basis]{
		\begin{tabular}{|*{2}{c|}}
			\hline
			$O_{2B}$  &  $\frac{g_{Y}^2}{160 \pi ^2 m_{\Delta }^2}$  \\
			\hline
			$O_{2W}$  &  $\frac{g_{W}^2}{240 \pi ^2 m_{\Delta }^2}$  \\
			\hline
			$O_{3W}$  &  $\frac{g_{W}^2}{240 \pi ^2 m_{\Delta }^2}$  \\
			\hline
			$O_6$  &  $-\frac{\zeta _1 \mu ^2}{m_{\Delta }^4}-\frac{\zeta _2 \mu ^2}{4 m_{\Delta }^4}-\frac{\zeta _1^3}{4 \pi ^2 m_{\Delta }^2}-\frac{\zeta _2^2 \zeta _1}{32 \pi ^2 m_{\Delta }^2}$  \\
			\hline
			$O_{\text{BB}}$  &  $\frac{\zeta _1}{32 \pi ^2 m_{\Delta }^2}$  \\
			\hline
			$O_H$  &  $\frac{\zeta _1^2}{8 \pi ^2 m_{\Delta }^2}+\frac{\mu ^2}{2 m_{\Delta }^4}$  \\
			\hline
			$O_R$  &  $\frac{\zeta _2^2}{96 \pi ^2 m_{\Delta }^2}+\frac{\mu ^2}{m_{\Delta }^4}$  \\
			\hline
			$O_T$  &  $\frac{\zeta _2^2}{192 \pi ^2 m_{\Delta }^2}-\frac{\mu ^2}{2 m_{\Delta }^4}$  \\
			\hline
			$O_{\text{WB}}$  &  $-\frac{\zeta _2}{96 \pi ^2 m_{\Delta }^2}$  \\
			\hline
			$O_{\text{WW}}$  &  $\frac{\zeta _1}{48 \pi ^2 m_{\Delta }^2}$  \\
			\hline
		\end{tabular}
			}
	\subfloat[``Warsaw'' basis]{
		\begin{tabular}{|*{2}{c|}}
			\hline
			$Q_H$  &  $-\frac{\zeta _1 \mu ^2}{m_{\Delta }^4}-\frac{\zeta _2 \mu \
				^2}{4 m_{\Delta }^4}-\frac{\zeta _1^3}{4 \pi ^2 m_{\Delta }^2}-\frac{\
				\zeta _2^2 \zeta _1}{32 \pi ^2 m_{\Delta }^2}$  \\
			\hline
			$Q_{H\square }$  &  $\frac{\zeta _2^2}{192 \pi ^2 m_{\Delta \
				}^2}+\frac{\mu ^2}{2 m_{\Delta }^4}$  \\
			\hline
			$Q_{\text{HD}}$  &  $\frac{\zeta _1^2}{4 \pi ^2 m_{\Delta \
				}^2}+\frac{\zeta _2^2}{96 \pi ^2 m_{\Delta }^2}-\frac{2 \mu \
				^2}{m_{\Delta }^4}$  \\
			\hline
			$Q_{\text{HW}}$  &  $\frac{\zeta _1 g_{W}^2}{48 \pi ^2 m_{\Delta \
				}^2}$  \\
			\hline
			$Q_{\text{HWB}}$  &  $-\frac{\zeta _2 g_{W} g_{Y}}{48 \pi ^2 \
				m_{\Delta }^2}$  \\
			\hline
			$Q_{\text{ll}}$  &  $\frac{\text{y$^2_\Delta $}}{4 m_{\Delta }^2}$  \
			\\
			\hline
			$Q_W$  &  $\frac{g_{W}^3}{1440 \pi ^2 m_{\Delta }^2}$  \\
			\hline
		\end{tabular}
	}\\
\subfloat[``Dimension-5" basis.]{
		\begin{tabular}{|c|c|}
			\hline
			Dimension-5 operator & Wilson Coefficient \\
			\hline
			$llHH$ & $\frac{y_{\Delta}^{2}}{m_{\Delta}}$ \\
			\hline
		\end{tabular}
	}
\end{table*}

	
	\item {\underline{Electro-weak $SU(2)_L$ Quartet Scalar with $Y=3/2$}}
{\small
	\begin{align}
	\label{lbsmQS}
	\mathcal{L}_{BSM}  & = \mathcal{L}_{SM} \ + (\mathcal{D}_{\mu} \Theta )^{\dagger} (\mathcal{D}^{\mu} \Theta ) -  m_{\Theta}^{2} \ |\Theta|^{2} - \kappa ( \Theta^{\dagger} B + h.c ), 
	\end{align}
}
	where the components of SM Higgs  H = $(H_{1} , H_{2})^{T}$ constitute $B$ as:
	\[
	B =
	\begin{bmatrix}
	H_{1}^{3} \\
	\sqrt{3} H_{1}^{2} H_{2}  \\
	\sqrt{3} H_{1} H_{2}^{2}  \\
	H_{2}^{3}
	\end{bmatrix}.
	\]
	
	Here, the heavy field is $\Theta$. The internal quantum numbers and its other required properties are given in Table~\ref{allbsmfields}. Once the $\Theta$ is integrated out using \mmaInlineCell{Input}{\mmaDef{CoDEx}}, the effective operators up to dimension-6 for all bases are generated which are listed in Table~\ref{tab:QS}.

\begin{table*}[!hbt]
	\caption{Effective operators and Wilson Coefficients in for Quartet Scalar  model.}
	\label{tab:QS}
	\centering
	\renewcommand{\arraystretch}{3.5}
	\subfloat[``SILH'' basis]{
		\begin{tabular}{|c|c|}
	\hline
	$O_{2B}$  &  $\frac{3 g_Y^2}{160 \pi ^2 m_{\Theta}^2}$  \\
	\hline
	$O_{2W}$  &  $\frac{g_W^2}{96 \pi ^2 m_{\Theta}^2}$  \\
	\hline
	$O_{3W}$  &  $\frac{g_W^2}{96 \pi ^2 m_{\Theta}^2}$  \\
	\hline
	$O_6$  &  $\frac{\kappa^2}{m_{\Theta}^2}$  \\
	\hline
\end{tabular}		
	}
	\subfloat[``Warsaw'' basis]{
		\begin{tabular}{|c|c|}
	\hline
	$Q_H$  &  $\frac{\kappa^2}{m_{\Theta}^2}$  \\
	\hline
	$Q_W$  &  $\frac{g_W^3}{576 \pi ^2 m_{\Theta}^2}$  \\
	\hline		
\end{tabular}		
	}
\end{table*}

	
	\item {\underline{$SU(2)_L$ singlet Heavy Right-handed neutrino}}
{\small	
	\begin{align}
	\label{lbsmT1SS}
	\mathcal{L}_{BSM} & =  \mathcal{L}_{SM} + \bar{\psi} ( i \slashed{\partial} - m_{\psi} ) \psi + [ y_{\psi} \bar{L} \tilde{H} \psi + h.c. ].
	\end{align}
}
	Here, the heavy field is $\psi$. The internal quantum numbers and its other required properties are given in Table~\ref{allbsmfields}. Once the $\psi$ is integrated out using \mmaInlineCell{Input}{\mmaDef{CoDEx}}, the dimension-5 effective operator is generated and is listed in Tables~\ref{tab:type-I-seesaw}.
	
	\begin{table}[!hbt]
		\caption{Type-I seesaw.}
		\label{tab:type-I-seesaw}
		\centering
		\renewcommand{\arraystretch}{2.5}
		\begin{tabular}{|c|c|}
			\hline
			Dimension-5 operators & Wilson Coefficient \\
			\hline
			$llHH$ & $\frac{y_{\psi}^{2}}{m_{\psi}}$ \\
			\hline
		\end{tabular}
	\end{table}
	
	
	\item \underline{{$SU(2)_L$ Real triplet Heavy fermion with $Y=0$}}
{\small	
	\begin{align}
	\label{lbsmT3SS}
	\mathcal{L}_{BSM} & =  \mathcal{L}_{SM} + \bar{\Sigma} ( i \slashed{\mathcal{D}} - m_{\Sigma} ) \bar{\Sigma} + [ y_{\Sigma} \bar{L} \Sigma \tilde{H} + h.c. ].
	\end{align}
}
	Here, the heavy field is $\Sigma$. The internal quantum numbers and its other required properties are given in Table~\ref{allbsmfields}. Once the $\Sigma$ is integrated out using \mmaInlineCell{Input}{\mmaDef{CoDEx}}, the dimension-5 effective operator is generated and is listed in Tables~\ref{tab:type-III-seesaw}.
	
	\begin{table}[!hbt]
		\caption{Type-III seesaw.}
		\label{tab:type-III-seesaw}
		\centering
		\renewcommand{\arraystretch}{2.5}
		\begin{tabular}{|c|c|}
			\hline
			Dimension-5 operators & Wilson Coefficient \\
			\hline
			$llHH$ & $\frac{y_{\Sigma}^{2}}{m_{\Sigma}}$ \\
			\hline
		\end{tabular}
	\end{table}
	
	
	\item {\underline{Heavy $SU(2)$ gauge boson}}
{\small
	\begin{align}
	\label{lbsmGB}
	\mathcal{L}_{BSM}  =  \mathcal{L}_{SM} &+ \frac{1}{2} \text{Q}_{\mu}^{a}  \left\lbrace  \mathcal{D}^{2} g^{\mu \nu} + m_{\text{Q}}^{2} g^{\mu \nu} + 2 [ \mathcal{D}^{\mu} , \mathcal{D}^{\nu} ] \right\rbrace^{ab} \text{Q}^{b}_{\nu} \nonumber \\
	& + \text{Q}_{\mu}^{a} \left(  \frac{g_{1}^{4}}{4(g_{1}^{4}+g_{2}^{4})} |H|^{2} g^{\mu \nu} \delta^{ab} \right)  \text{Q}^{b}_{\nu} + \frac{g_{1}^{2}}{\sqrt{g_{1}^{2}+g_{2}^{2}}} \text{Q}_{\mu}^{a} \mathcal{D}_{\nu} W^{a , \nu \mu} + \ldots
	\end{align}
}
	
	Here, the heavy field is $Q_\mu$. The internal quantum numbers and its other required properties are given in Table~\ref{allbsmfields}. Once the $Q_\mu$ is integrated out using \mmaInlineCell{Input}{\mmaDef{CoDEx}}, the effective operators upto dimension-6 for both \mmaInlineCell{Code}{"SILH"} and \mmaInlineCell{Code}{"Warsaw"} bases are generated which are listed in Table~\ref{tab:GB}.

\begin{table*}[!hbt]
	\caption{Effective operators and Wilson Coefficients for Exotic $SU(2)$ gauge boson model.}
	\label{tab:GB}
	\small
	\centering
	\renewcommand{\arraystretch}{2.5}
	\subfloat[``SILH'' basis]{
		\begin{tabular}{|c|c|}
	\hline
	$O_{2W}$  &  $\frac{g_1^4}{m_Q^2 \left(g_1^2+g_2^2\right)}-\frac{37 g_W^2}{480 \pi ^2 m_Q^2}$  \\
	\hline
	$O_{3W}$  &  $\frac{g_W^2}{160 \pi ^2 m_Q^2}$  \\
	\hline
	$O_6$  &  $-\frac{g_1^{12}}{576 \pi ^2 m_Q^2 \left(g_1^2+g_2^2\right)^3}$  \\
	\hline
	$O_H$  &  $\frac{g_1^8}{64 \pi ^2 m_Q^2 \left(g_1^2+g_2^2\right)^2}$  \\
	\hline
	$O_{\text{WW}}$  &  $\frac{g_1^4}{48 \pi ^2 m_Q^2 \left(g_1^2+g_2^2\right)}$  \\
	\hline
\end{tabular}
	}
	\subfloat[``Warsaw'' basis]{
		\begin{tabular}{|c|c|}
	\hline
	$Q_H$  &  $-\frac{g_1^{12}}{576 \pi ^2 m_Q^2 \left(g_1^2+g_2^2\right)^3}$  \\
	\hline
	$Q_{\text{HD}}$  &  $\frac{g_1^8}{32 \pi ^2 m_Q^2 \left(g_1^2+g_2^2\right)^2}$  \\
	\hline
	$Q_{\text{HW}}$  &  $\frac{g_1^4 g_W^2}{48 \pi ^2 m_Q^2 \left(g_1^2+g_2^2\right)}$  \\
	\hline
	$Q_W$  &  $\frac{g_W^3}{960 \pi ^2 m_Q^2}$  \\
	\hline
\end{tabular}
	}
\end{table*}

	
	\item {\underline{Two Higgs Doublet Model (2HDM)}}
{\small	
	\begin{align}
	\label{lbsm2HDM}
	\mathcal{L}_{BSM} =  \mathcal{L}_{SM} \ &+ \ |\mathcal{D}_{\mu} \ \varphi |^{2} \ - \ m_{\varphi}^{2} \ |\varphi|^{2} - \ \frac{\lambda_{\varphi}}{4} |\varphi|^{4} + ( \eta_{H} |\tilde{H} |^{2} + \eta_{\varphi} |\varphi|^{2}) (\tilde{H}^{\dagger} \varphi + \varphi^{\dagger} \tilde{H}) \nonumber \\ 
	& - \lambda_{1} |\tilde{H}|^{2} |\varphi|^{2} - \lambda_{2} |\tilde{H}^{\dagger} \varphi |^{2} - \lambda_{3} \left[ (\tilde{H}^{\dagger} \varphi)^{2} + (\varphi^{\dagger} \tilde{H})^{2} \right] 
	\end{align}
}
	Here, the heavy field is $\varphi$. The internal quantum numbers and its other required properties are given in Table~\ref{allbsmfields}. Once the $\varphi$ is integrated out using \mmaInlineCell{Input}{\mmaDef{CoDEx}}, the effective operators upto dimension-6 for both \mmaInlineCell{Code}{"SILH"} and \mmaInlineCell{Code}{"Warsaw"} bases are generated which are listed in Table~\ref{tab:2HDM}.
	
\begin{table*}[!hbt]
	\caption{Effective operators and Wilson Coefficients for Two Higgs Doublet Model (2HDM).}
	\label{tab:2HDM}
	\small
	\centering
	\renewcommand{\arraystretch}{2}
	\subfloat[``SILH'' basis]{
		\begin{tabular}{|c|c|}
	\hline
	$O_{H}$ & $-\frac{3 \eta_{H} \eta_{\varphi}}{8 \pi^{2} m_{\varphi}^2} + \frac{\lambda_{1} \lambda_{2}}{48 \pi^{2} m_{\varphi}^{2}} + \frac{\lambda_{1}^{2}}{48 \pi^{2} m_{\varphi}^2} + \frac{\lambda_{2}^{2}}{192 \pi^{2} m_{\varphi}^2} + \frac{\lambda_{3}^{2}}{48 \pi^{2} m_{\varphi}^2}$ \\
	\hline
	$O_{T}$ & $ \frac{\lambda_{2}^{2}}{192 \pi^{2} m_{\varphi}^2} - \frac{\lambda_{3}^{2}}{48 \pi^{2} m_{\varphi}^2}$\\
	\hline
	$O_{R}$ & $ - \frac{3 \eta_{H} \eta_{\varphi}}{8 \pi^{2} m_{\varphi}^2} + \frac{\lambda_{2}^{2}}{96 \pi^{2} m_{\varphi}^2} + \frac{\lambda_{3}^{2}}{24 \pi^{2} m_{\varphi}^2}$ \\
	\hline
	$O_{6}$ & $ \frac{\eta_H^2}{m_{\varphi}^{2}}+ \frac{3 \eta_{H} \eta_{\varphi} \lambda_{1}}{8 \pi^{2} m_{\varphi}^2 } - \frac{\lambda_1^3}{48 \pi^{2} m_{\varphi}^2} + \frac{3 \eta_{H} \eta_{\varphi} \lambda_{2}}{8 \pi^{2} m_{\varphi}^2} - \frac{\lambda_{1}^{2} \lambda_{2}}{32 \pi^{2} m_{\varphi}^2}$ \\
	&$ - \frac{\lambda_{1}\lambda_{2}^{2}}{32 \pi^{2} m_{\varphi}^2} - \frac{\lambda_{2}^{3}}{96 \pi^{2} m_{\varphi}^2} - \frac{\lambda_{1} \lambda_{3}^{2}}{8 \pi^{2} m_{\varphi}^2} - \frac{\lambda_{2} \lambda_{3}^{2}}{8 \pi^{2} m_{\varphi}^{2}} + \frac{3 \eta_{H}^{2} \lambda_{\varphi}}{32 \pi^{2} m_{\varphi}^2}$\\
	\hline
	$O_{WW}$ & $\frac{ 2 \lambda_{1} + \lambda_{2}}{768 \pi^{2} m_{\varphi}^{2} }$ \\
	\hline
	$O_{2W}$ & $\frac{g_{W}^{2}}{960 \pi^{2} m_{\varphi}^{2} }$ \\
	\hline
	$O_{3W}$ & $\frac{g_{W}^{2}}{960 \pi^{2} m_{\varphi}^{2} }$ \\
	\hline
	$O_{WB}$ & $\frac{\lambda_{2}}{384 \pi^{2} m_{\varphi}^{2}}$ \\
	\hline
	$O_{BB}$ & $\frac{ 2 \lambda_{1} + \lambda_{2}}{768 \pi^{2} m_{\varphi}^{2} }$ \\
	\hline
	$O_{2B}$ & $\frac{g_{Y}^{2}}{960 \pi^{2} m_{\varphi}^{2}}$ \\
	\hline
\end{tabular}
}
	\subfloat[``Warsaw'' basis]{
		\begin{tabular}{|c|c|}
	\hline
	$Q_H$  &  $\frac{3 \eta_H^2 \lambda_\varphi}{32 \pi ^2 m_{\varphi}^{2}}+\frac{\eta_H^2}{m_{\varphi}^{2}}+\frac{3 \eta_H \eta_\phi \lambda_1}{8 \pi ^2 m_{\varphi}^{2}}+\frac{3 \eta_H \eta_\phi \lambda_2}{8 \pi ^2 m_{\varphi}^{2}}-\frac{\lambda_1^3}{48 \pi ^2 m_{\varphi}^{2}}$\\
	&-$\frac{\lambda_1^2 \lambda_2}{32 \pi ^2 m_{\varphi}^{2}}-\frac{\lambda_1 \lambda_2^2}{32 \pi ^2 m_{\varphi}^{2}}-\frac{\lambda_1 \lambda_3^2}{8 \pi ^2 m_{\varphi}^{2}}-\frac{\lambda_2^3}{96 \pi ^2 m_{\varphi}^{2}}-\frac{\lambda_2 \lambda_3^2}{8 \pi ^2 m_{\varphi}^{2}}$  \\
	\hline
	$Q_{H\square }$  &  $-\frac{3 \eta_H \eta_\varphi}{16 \pi ^2 m_{\varphi}^{2}}+\frac{\lambda_2^2}{192 \pi ^2 m_{\varphi}^{2}}+\frac{\lambda_3^2}{48 \pi ^2 m_{\varphi}^{2}}$  \\
	\hline
	$Q_{\text{HD}}$  &  $\frac{9 \eta_H \eta_\varphi}{8 \pi ^2 m_{\varphi}^{2}}+\frac{\lambda_1^2}{24 \pi ^2 m_{\varphi}^{2}}+\frac{\lambda_1 \lambda_2}{24 \pi ^2 m_{\varphi}^{2}}+\frac{\lambda_2^2}{48 \pi ^2 m_{\varphi}^{2}}+\frac{\lambda_3^2}{12 \pi ^2 m_{\varphi}^{2}}$  \\
	\hline
	$Q_{\text{HW}}$  &  $\frac{g_W^2 \lambda_1}{384 \pi ^2 m_{\varphi}^{2}}+\frac{g_W^2 \lambda_2}{768 \pi ^2 m_{\varphi}^{2}}$  \\
	\hline
	$Q_{\text{HWB}}$  &  $\frac{g_W g_Y \lambda_2}{192 \pi ^2 m_{\varphi}^{2}}$  \\
	\hline
	$Q_W$  &  $\frac{g_W^3}{5760 \pi ^2 m_{\varphi}^{2}}$  \\
	\hline
\end{tabular}
}
\end{table*}
	
	\item {\underline{Exotic U(1) gauge boson}}
{\small	
	\begin{align}
	\label{U1GB}
	\mathcal{L}_{BSM} = -\frac{1}{4} K_{\mu \nu} K^{\mu \nu} + \frac{1}{2} m_{K}^{2} K_{\mu} K^{\mu} - \frac{k}{2} B^{\mu \nu} K_{\mu \nu}.
	\end{align}
}
	Here, $K_{\mu \nu}$ is the field strength corresponding to the the heavy field $K_\mu$. The internal quantum numbers and its other required properties are given in Table~\ref{allbsmfields}. Once the $K_\mu$ is integrated out using \mmaInlineCell{Input}{\mmaDef{CoDEx}}, the effective operators of dimension-6 for \mmaInlineCell{Code}{"SILH"} basis are generated which are listed in Table~\ref{yew1}.
	
	\begin{table*}[!hbt]
		\caption{Effective operators and Wilson Coefficients in ``SILH'' basis for Exotic $U(1)$ gauge bosons.}
				\label{yew1}
		\centering
		\renewcommand*{\arraystretch}{2.5}
		\begin{tabular}{|c|c|}
			\hline
			Dimension-6 operators & Wilson Coefficient \\
			\hline
			$O_{2B}$  &  $\frac{k^2}{m_{K}^2}$  \\
			\hline
		\end{tabular}
	\end{table*}
	
\end{enumerate}

\section{Examples:  Multiple heavy BSM fields}

\begin{enumerate}
	
	\item{\underline{Electro-weak $SU(2)_L$ Real Singlet ($\phi$)+ Triplet ($\Phi$) Scalars with $Y=0$}}
{\small	
	\begin{align}
	\label{lbsmSINGTRIP}
	\mathcal{L}_{BSM} & =   \mathcal{L}_{SM} \ + \ \frac{1}{2} \ (\mathcal{\partial}_{\mu} \phi )^{2} \ - \ \frac{1}{2} \ m_{\phi}^{2} \ \phi^{2} - \mu_{\phi } |H|^{2} \phi  - \frac{1}{2} \kappa_{\phi} |H|^{2}  \phi^{2} - \frac{1}{3!} \ \mu \phi^{3}  - \frac{1}{4!} \ \lambda_{\phi} \phi^{4} \nonumber \\
	& + \ \frac{1}{2} \ (\mathcal{D}_{\mu} \Phi )^{2} \ - \ \frac{1}{2} \ m_{\Phi}^{2} \ \Phi^{a} \ \Phi^{a} + 2 \ \kappa_{\Phi} \ H^{\dagger} \tau^{a} H \ \Phi^{a} - \ \eta \ |H|^{2} \ \Phi^{a} \ \Phi^{a} - \frac{1}{4} \ \lambda_{\Phi} \ (\Phi^{a} \Phi^{a})^{2} + 2 \beta \ (H^{\dagger} \tau^{a} H) \ \Phi^{a} \ \phi.
	\end{align}
}
	Here, the heavy fields $\Phi$ and $\phi$ are the same as mentioned in earlier section. The interaction term among these heavy field is given as $2 \beta \ (H^{\dagger} \tau^{a} H) \ \Phi^{a}$.  Once the $\Phi$  and $\phi$ are integrated out using \mmaInlineCell{Input}{\mmaDef{CoDEx}}, the effective operators upto dimension-6 for both \mmaInlineCell{Code}{"SILH"} and \mmaInlineCell{Code}{"Warsaw"} bases are generated which are listed in Table~\ref{tab:ST}.
	
\begin{table*}[!hbt]
	\caption{Effective operators and Wilson Coefficients in for Real Singlet and Triplet (with Y=0) models. These two fields are degenerate in mass, i.e., $m_{\Phi }=m_{\phi }$.}
	\label{tab:ST}
	\small
	\centering
	\renewcommand{\arraystretch}{2}
	\subfloat[``SILH'' basis]{
		\begin{tabular}{|*{2}{c|}}
	\hline
	$O_{2W}$  &  $\frac{g_{W}^2}{480 \pi ^2 m_{\Phi }^2}$  \\
	\hline
	$O_{3W}$  &  $\frac{g_{W}^2}{480 \pi ^2 m_{\Phi }^2}$  \\
	\hline
	$O_6$  &  $-\frac{\beta ^2 \eta }{32 \pi ^2 m_{\Phi }^2}-\frac{\beta ^2 \kappa _{\phi }}{64 \pi ^2 m_{\Phi }^2}+\frac{\beta ^2 \mu  \mu _{\phi }}{64 \pi ^2 m_{\Phi }^2 m_{\phi }^2}-$ \\
	& $\frac{\beta  \kappa _{\Phi } \mu _{\phi }}{m_{\Phi }^2 m_{\phi }^2}-\frac{\eta ^3}{8 \pi ^2 m_{\Phi }^2}-\frac{5 \eta  \kappa _{\Phi }^2 \lambda _{\Phi }}{8 \pi ^2 m_{\Phi }^4}-\frac{\eta  \kappa _{\Phi }^2}{m_{\Phi }^4}-$ \\
	& $\frac{\kappa _{\phi } \lambda _{\phi } \mu _{\phi }^2 m_{\Phi }^2}{32 \pi ^2 m_{\phi }^6}-\frac{\mu ^2 \kappa _{\phi } \mu _{\phi }^2}{64 \pi ^2 m_{\Phi }^2 m_{\phi }^4}+\frac{\mu  \kappa _{\phi }^2 \mu _{\phi }}{64 \pi ^2 m_{\Phi }^2 m_{\phi }^2}-$ \\
	& $\frac{\kappa _{\phi } \mu _{\phi }^2}{2 m_{\phi }^4}-\frac{\kappa _{\phi }^3}{192 \pi ^2 m_{\Phi }^2}+\frac{\mu ^3 \mu _{\phi }^3}{192 \pi ^2 m_{\Phi }^2 m_{\phi }^6}+\frac{\mu  \mu _{\phi }^3}{6 m_{\phi }^6}$  \\
	\hline
	$O_H$  &  $\frac{\eta ^2}{16 \pi ^2 m_{\Phi }^2}+\frac{\kappa _{\phi }^2}{192 \pi ^2 m_{\Phi }^2}+\frac{\lambda _{\phi } \mu _{\phi }^2 m_{\Phi }^2}{16 \pi ^2 m_{\phi }^6}+\frac{\mu _{\phi }^2}{m_{\phi }^4}$  \\
	\hline
	$O_R$  &  $\frac{\beta ^2}{48 \pi ^2 m_{\Phi }^2}+\frac{5 \kappa _{\Phi }^2 \lambda _{\Phi }}{4 \pi ^2 m_{\Phi }^4}+\frac{2 \kappa _{\Phi }^2}{m_{\Phi }^4}$  \\
	\hline
	$O_T$  &  $\frac{\beta ^2}{96 \pi ^2 m_{\Phi }^2}+\frac{5 \kappa _{\Phi }^2 \lambda _{\Phi }}{8 \pi ^2 m_{\Phi }^4}+\frac{\kappa _{\Phi }^2}{m_{\Phi }^4}$  \\
	\hline
	$O_{\text{WW}}$  &  $\frac{\eta }{96 \pi ^2 m_{\Phi }^2}$  \\
	\hline
\end{tabular}
}
	\subfloat[``Warsaw'' basis]{
\begin{tabular}{|*{2}{c|}}
	\hline
	$Q_H$  &  $-\frac{\beta ^2 \eta }{32 \pi ^2 m_{\Phi }^2}-\frac{\beta ^2 \kappa _{\phi }}{64 \pi ^2 m_{\Phi }^2}+\frac{\beta ^2 \mu  \mu _{\phi }}{64 \pi ^2 m_{\Phi }^2 m_{\phi }^2}-\frac{\beta  \kappa _{\Phi } \mu _{\phi }}{m_{\Phi }^2 m_{\phi }^2} $ \\ 
	& $ -\frac{\eta ^3}{8 \pi ^2 m_{\Phi }^2}-\frac{5 \eta  \kappa _{\Phi }^2 \lambda _{\Phi }}{8 \pi ^2 m_{\Phi }^4}-\frac{\eta  \kappa _{\Phi }^2}{m_{\Phi }^4}-\frac{\kappa _{\phi } \lambda _{\phi } \mu _{\phi }^2 m_{\Phi }^2}{32 \pi ^2 m_{\phi }^6} $ \\ 
	& $ -\frac{\mu ^2 \kappa _{\phi } \mu _{\phi }^2}{64 \pi ^2 m_{\Phi }^2 m_{\phi }^4}+\frac{\mu  \kappa _{\phi }^2 \mu _{\phi }}{64 \pi ^2 m_{\Phi }^2 m_{\phi }^2}-\frac{\kappa _{\phi } \mu _{\phi }^2}{2 m_{\phi }^4}-\frac{\kappa _{\phi }^3}{192 \pi ^2 m_{\Phi }^2}+\frac{\mu ^3 \mu _{\phi }^3}{192 \pi ^2 m_{\Phi }^2 m_{\phi }^6}+\frac{\mu  \mu _{\phi }^3}{6 m_{\phi }^6}$  \\
	\hline
	$Q_{H\square }$  &  $\frac{\beta ^2}{96 \pi ^2 m_{\Phi }^2}+\frac{5 \kappa _{\Phi }^2 \lambda _{\Phi }}{8 \pi ^2 m_{\Phi }^4}+\frac{\kappa _{\Phi }^2}{m_{\Phi }^4}$  \\
	\hline
	$Q_{\text{HD}}$  &  $\frac{\beta ^2}{48 \pi ^2 m_{\Phi }^2}+\frac{\eta ^2}{8 \pi ^2 m_{\Phi }^2}-\frac{5 \kappa _{\Phi }^2 \lambda _{\Phi }}{4 \pi ^2 m_{\Phi }^4}-\frac{2 \kappa _{\Phi }^2}{m_{\Phi }^4}+\frac{\kappa _{\phi }^2}{96 \pi ^2 m_{\Phi }^2}-\frac{\lambda _{\phi } \mu _{\phi }^2 m_{\Phi }^2}{8 \pi ^2 m_{\phi }^6}-\frac{2 \mu _{\phi }^2}{m_{\phi }^4}$  \\
	\hline
	$Q_{\text{HW}}$  &  $\frac{\eta ~ g_W^2}{96 \pi ^2 m_{\Phi }^2}$  \\
	\hline
	$Q_W$  &  $\frac{g_W^3}{2880 \pi ^2 m_{\Phi }^2}$  \\
	\hline
\end{tabular}

}
\end{table*}

	
	\item{\underline{Electro-weak Complex triplet $(\Delta)$and Complex doublet ($\varphi$) with $Y=1$ models}}

	Complex doublet and complex triplet Lagrangian :-
{\small 	
	\begin{align}
	\label{lbsmCTSnCDS}
	\mathcal{L}_{BSM} = \ & \ \mathcal{L}_{SM} \ + Tr [ (\mathcal{D}_{\mu} \Delta)^{\dagger} (\mathcal{D}^{\mu} \Delta ) ] - m_{\Delta}^{2} Tr [ \Delta^{\dagger} \Delta ] + \mathcal{L}_{Y} - V(H,\Delta) + \mathcal{L}_{2} + \mathcal{L}_{int} \nonumber 
	\end{align}
}
	where,
{\small 
	\begin{align}
	V(H,\Delta) =& \zeta_{1} (H^{\dagger} H) Tr[ \Delta^{\dagger} \Delta ] + \zeta_{2} (H^{\dagger} \tau^{i} H) Tr[ \Delta^{\dagger} \tau^{i} \Delta ] + [ \mu ( H^{T} i \sigma^{2} \Delta^{\dagger} H ) + h.c.], \\
	\mathcal{L}_{Y} =& y_{\Delta} L^{T} C i \tau^2 \Delta L + h.c.   ,\\
	\mathcal{L}_{2} =& \ |\mathcal{D}_{\mu} \ \varphi |^{2} \ - \ m_{\varphi}^{2} \ |\varphi|^{2} - \ \frac{\lambda_{\varphi}}{4} |\varphi|^{4} + ( \eta_{H} |\tilde{H} |^{2} + \eta_{\varphi} |\varphi|^{2}) (\tilde{H}^{\dagger} \varphi + \varphi^{\dagger} \tilde{H}) - \lambda_{1} |\tilde{H}|^{2} |\varphi|^{2} \nonumber \\ 
	& - \lambda_{2} |\tilde{H}^{\dagger} \varphi |^{2} - \lambda_{3} \left[ (\tilde{H}^{\dagger} \varphi)^{2} + (\varphi^{\dagger} \tilde{H})^{2} \right] ,\\
	\text{and,}~~~~~\mathcal{L}_{int} &= \mu_{1} H^{\dagger} \Delta \varphi + h.c.
	\end{align}
}	
	
	Here, the heavy fields $\Delta$ and $\varphi$ are the same as mentioned in earlier section. The interaction term among these heavy field is given in $\mathcal{L}_{int}$.  Once the $\Delta$  and $\varphi$ are integrated out using \mmaInlineCell{Input}{\mmaDef{CoDEx}}, the effective operators upto dimension-6 for both \mmaInlineCell{Code}{"SILH"} and \mmaInlineCell{Code}{"Warsaw"} bases are generated which are listed in Table~\ref{tab:CDT}.

\begin{table*}[!hbt]
	\caption{Effective operators and Wilson Coefficients for Complex doublet and Complex Triplet models. These two fields are degenerate in mass, i.e., $m_{\Delta }=m_{\varphi }$.}
	\label{tab:CDT}
	\small
	\centering
	\renewcommand{\arraystretch}{2}
	\subfloat[``SILH'' basis]{
		\begin{tabular}{|*{2}{c|}}
	\hline
	$O_{2B}$  &  $\frac{7 g_Y^2}{960 \pi ^2 m_{\Delta }^2}$  \\
	\hline
	$O_{2W}$  &  $\frac{g_W^2}{192 \pi ^2 m_{\Delta }^2}$  \\
	\hline
	$O_{3W}$  &  $\frac{g_W^2}{192 \pi ^2 m_{\Delta }^2}$  \\
	\hline
	$O_6$  &  $-\frac{3 \zeta _1 \mu ^2 \lambda _{\varphi } m_{\Delta }^2}{16 \pi ^2 m_{\varphi }^6}+\frac{3 \zeta _2 \mu ^2 \lambda _{\varphi } m_{\Delta }^2}{64 \pi ^2 m_{\varphi }^6}+\frac{\zeta _1 \lambda _1 \mu _1^2}{16 \pi ^2 m_{\Delta }^4}+\frac{\zeta _1 \lambda _2 \mu _1^2}{24 \pi ^2 m_{\Delta }^4} $ \\
	& $ -\frac{\zeta _1 \mu ^2}{m_{\Delta }^4}-\frac{\zeta _2 \mu ^2}{4 m_{\Delta }^4}  -\frac{5 \zeta _1 \mu _1^4}{96 \pi ^2 m_{\Delta }^6}+\frac{\zeta _1^2 \mu _1^2}{8 \pi ^2 m_{\Delta }^4} $ \\
	& $ +\frac{\zeta _2^2 \mu _1^2}{192 \pi ^2 m_{\Delta }^4}-\frac{\zeta _1^3}{4 \pi ^2 m_{\Delta }^2}-\frac{\zeta _1 \zeta _2^2}{32 \pi ^2 m_{\Delta }^2}-\frac{9 \mu ^2 \mu _1^2 \lambda _{\varphi }}{256 \pi ^2 m_{\Delta }^2 m_{\varphi }^4}$ \\
	& $ -\frac{5 \lambda _1 \mu _1^4}{192 \pi ^2 m_{\Delta }^6}-\frac{\lambda _2 \mu _1^4}{48 \pi ^2 m_{\Delta }^6} +\frac{\lambda _1^2 \mu _1^2}{32 \pi ^2 m_{\Delta }^4}+\frac{\lambda _2^2 \mu _1^2}{48 \pi ^2 m_{\Delta }^4}$ \\
	& $ +\frac{\lambda _3^2 \mu _1^2}{12 \pi ^2 m_{\Delta }^4}+\frac{\lambda _1 \lambda _2 \mu _1^2}{24 \pi ^2 m_{\Delta }^4}-\frac{\lambda _1^3}{48 \pi ^2 m_{\Delta }^2}-\frac{\lambda _2^3}{96 \pi ^2 m_{\Delta }^2} $ \\
	& $ -\frac{\lambda _1 \lambda _2^2}{32 \pi ^2 m_{\Delta }^2}  -\frac{\lambda _1 \lambda _3^2}{8 \pi ^2 m_{\Delta }^2}-\frac{\lambda _2 \lambda _3^2}{8 \pi ^2 m_{\Delta }^2}$ \\
	& $ -\frac{\lambda _1^2 \lambda _2}{32 \pi ^2 m_{\Delta }^2}+\frac{3 \mu _1^6}{320 \pi ^2 m_{\Delta }^8}+\frac{\text{$\eta $H}^2}{m_{\varphi }^2}$  \\
	\hline
	$O_B$  &  $\frac{3 \mu _1^2}{640 \pi ^2 m_{\Delta }^4}$  \\
	\hline
	$O_{\text{BB}}$  &  $\frac{\zeta _1}{32 \pi ^2 m_{\Delta }^2}+\frac{\lambda _1}{384 \pi ^2 m_{\Delta }^2}+\frac{\lambda _2}{768 \pi ^2 m_{\Delta }^2}-\frac{7 \mu _1^2}{2560 \pi ^2 m_{\Delta }^4}$  \\
	\hline
	$O_D$  &  $\frac{\mu _1^2}{320 \pi ^2 m_{\Delta }^4}$  \\
	\hline
	$O_H$  &  $-\frac{\zeta _1 \mu _1^2}{16 \pi ^2 m_{\Delta }^4}+\frac{\zeta _1^2}{8 \pi ^2 m_{\Delta }^2}-\frac{\lambda _1 \mu _1^2}{32 \pi ^2 m_{\Delta }^4}-\frac{7 \lambda _2 \mu _1^2}{384 \pi ^2 m_{\Delta }^4}+\frac{\lambda _1^2}{48 \pi ^2 m_{\Delta }^2} $ \\
	& $ +\frac{\lambda _2^2}{192 \pi ^2 m_{\Delta }^2}+\frac{\lambda _3^2}{48 \pi ^2 m_{\Delta }^2}+\frac{\lambda _1 \lambda _2}{48 \pi ^2 m_{\Delta }^2}+\frac{\mu ^2}{2 m_{\Delta }^4}+\frac{7 \mu _1^4}{384 \pi ^2 m_{\Delta }^6}$  \\
	\hline
	$O_R$  &  $-\frac{\zeta _1 \mu _1^2}{32 \pi ^2 m_{\Delta }^4}+\frac{\zeta _2^2}{96 \pi ^2 m_{\Delta }^2}-\frac{\lambda _1 \mu _1^2}{64 \pi ^2 m_{\Delta }^4}-\frac{\lambda _2 \mu _1^2}{64 \pi ^2 m_{\Delta }^4}+\frac{\lambda _2^2}{96 \pi ^2 m_{\Delta }^2} $ \\
	& $ +\frac{\lambda _3^2}{24 \pi ^2 m_{\Delta }^2}+\frac{\mu ^2}{m_{\Delta }^4}+\frac{\mu _1^4}{64 \pi ^2 m_{\Delta }^6}$  \\
	\hline
	$O_T$  &  $\frac{\zeta _2^2}{192 \pi ^2 m_{\Delta }^2}+\frac{\lambda _2 \mu _1^2}{384 \pi ^2 m_{\Delta }^4}+\frac{\lambda _2^2}{192 \pi ^2 m_{\Delta }^2} $ \\ 
	& $-\frac{\lambda _3^2}{48 \pi ^2 m_{\Delta }^2}-\frac{\mu ^2}{2 m_{\Delta }^4}+\frac{\mu _1^4}{384 \pi ^2 m_{\Delta }^6}$  \\
	\hline
	$O_W$  &  $\frac{\mu _1^2}{1920 \pi ^2 m_{\Delta }^4}$  \\
	\hline
	$O_{\text{WB}}$  &  $-\frac{\zeta _2}{96 \pi ^2 m_{\Delta }^2}+\frac{\lambda _2}{384 \pi ^2 m_{\Delta }^2}+\frac{\mu _1^2}{3840 \pi ^2 m_{\Delta }^4}$  \\
	\hline
	$O_{\text{WW}}$  &  $\frac{\zeta _1}{48 \pi ^2 m_{\Delta }^2}+\frac{\lambda _1}{384 \pi ^2 m_{\Delta }^2}+\frac{\lambda _2}{768 \pi ^2 m_{\Delta }^2}-\frac{11 \mu _1^2}{2560 \pi ^2 m_{\Delta }^4}$  \\
	\hline
\end{tabular}
	}
\subfloat[``Warsaw'' basis]{
		\begin{tabular}{|*{2}{c|}}
	\hline
	$Q_H$  &  $-\frac{3 \zeta _1 \mu ^2 \lambda _{\varphi } m_{\Delta }^2}{16 \pi ^2 m_{\varphi }^6}+\frac{3 \zeta _2 \mu ^2 \lambda _{\varphi } m_{\Delta }^2}{64 \pi ^2 m_{\varphi }^6}+\frac{\zeta _1 \lambda _1 \mu _1^2}{16 \pi ^2 m_{\Delta }^4} $ \\ 
	& $ +\frac{\zeta _1 \lambda _2 \mu _1^2}{24 \pi ^2 m_{\Delta }^4}-\frac{\zeta _1 \mu ^2}{m_{\Delta }^4} -\frac{\zeta _2 \mu ^2}{4 m_{\Delta }^4}-\frac{5 \zeta _1 \mu _1^4}{96 \pi ^2 m_{\Delta }^6}$ \\
	& $ +\frac{\zeta _1^2 \mu _1^2}{8 \pi ^2 m_{\Delta }^4}+\frac{\zeta _2^2 \mu _1^2}{192 \pi ^2 m_{\Delta }^4}-\frac{\zeta _1^3}{4 \pi ^2 m_{\Delta }^2}-\frac{\zeta _1 \zeta _2^2}{32 \pi ^2 m_{\Delta }^2} $ \\
	& $-\frac{9 \mu ^2 \mu _1^2 \lambda _{\varphi }}{256 \pi ^2 m_{\Delta }^2 m_{\varphi }^4}-\frac{5 \lambda _1 \mu _1^4}{192 \pi ^2 m_{\Delta }^6}-\frac{\lambda _2 \mu _1^4}{48 \pi ^2 m_{\Delta }^6}+\frac{\lambda _1^2 \mu _1^2}{32 \pi ^2 m_{\Delta }^4} $ \\ 
	& $ +\frac{\lambda _2^2 \mu _1^2}{48 \pi ^2 m_{\Delta }^4}+\frac{\lambda _3^2 \mu _1^2}{12 \pi ^2 m_{\Delta }^4} +\frac{\lambda _1 \lambda _2 \mu _1^2}{24 \pi ^2 m_{\Delta }^4}-\frac{\lambda _1^3}{48 \pi ^2 m_{\Delta }^2}$ \\
	& $ -\frac{\lambda _2^3}{96 \pi ^2 m_{\Delta }^2}-\frac{\lambda _1 \lambda _2^2}{32 \pi ^2 m_{\Delta }^2}-\frac{\lambda _1 \lambda _3^2}{8 \pi ^2 m_{\Delta }^2}-\frac{\lambda _2 \lambda _3^2}{8 \pi ^2 m_{\Delta }^2} $ \\
	& $ -\frac{\lambda _1^2 \lambda _2}{32 \pi ^2 m_{\Delta }^2}+\frac{3 \mu _1^6}{320 \pi ^2 m_{\Delta }^8}+\frac{\text{$\eta _{H}$}^2}{m_{\varphi }^2}$  \\
	\hline
	$Q_{\text{HB}}$  &  $\frac{\zeta _1 g_{Y}^2}{32 \pi ^2 m_{\Delta }^2}+\frac{g_{Y}^2 \lambda _1}{384 \pi ^2 m_{\Delta }^2}+\frac{g_{Y}^2 \lambda _2}{768 \pi ^2 m_{\Delta }^2}-\frac{7 g_{Y}^2 \mu _1^2}{2560 \pi ^2 m_{\Delta }^4}$  \\
	\hline
	$Q_{H\square }$  &  $-\frac{\zeta _1 \mu _1^2}{64 \pi ^2 m_{\Delta }^4}+\frac{\zeta _2^2}{192 \pi ^2 m_{\Delta }^2}-\frac{\lambda _1 \mu _1^2}{128 \pi ^2 m_{\Delta }^4}-\frac{\lambda _2 \mu _1^2}{128 \pi ^2 m_{\Delta }^4} $ \\
	& $ +\frac{\lambda _2^2}{192 \pi ^2 m_{\Delta }^2}+\frac{\lambda _3^2}{48 \pi ^2 m_{\Delta }^2}+\frac{\mu ^2}{2 m_{\Delta }^4}+\frac{\mu _1^4}{128 \pi ^2 m_{\Delta }^6}$  \\
	\hline
	$Q_{\text{HD}}$  &  $-\frac{5 \zeta _1 \mu _1^2}{32 \pi ^2 m_{\Delta }^4}+\frac{\zeta _1^2}{4 \pi ^2 m_{\Delta }^2}+\frac{\zeta _2^2}{96 \pi ^2 m_{\Delta }^2} $ \\ 
	& $ -\frac{5 \lambda _1 \mu _1^2}{64 \pi ^2 m_{\Delta }^4}-\frac{5 \lambda _2 \mu _1^2}{96 \pi ^2 m_{\Delta }^4} +\frac{\lambda _1^2}{24 \pi ^2 m_{\Delta }^2}+\frac{\lambda _2^2}{48 \pi ^2 m_{\Delta }^2}$ \\
	& $ +\frac{\lambda _3^2}{12 \pi ^2 m_{\Delta }^2}+\frac{\lambda _1 \lambda _2}{24 \pi ^2 m_{\Delta }^2}-\frac{2 \mu ^2}{m_{\Delta }^4}+\frac{5 \mu _1^4}{96 \pi ^2 m_{\Delta }^6}$  \\
	\hline
	$Q_{\text{HW}}$  &  $\frac{\zeta _1 g_{W}^2}{48 \pi ^2 m_{\Delta }^2}+\frac{g_{W}^2 \lambda _1}{384 \pi ^2 m_{\Delta }^2}+\frac{g_{W}^2 \lambda _2}{768 \pi ^2 m_{\Delta }^2}-\frac{37 g_{W}^2 \mu _1^2}{7680 \pi ^2 m_{\Delta }^4}$  \\
	\hline
	$Q_{\text{HWB}}$  &  $-\frac{\zeta _2 g_{W} g_{Y}}{48 \pi ^2 m_{\Delta }^2}+\frac{g_{W} g_{Y} \lambda _2}{192 \pi ^2 m_{\Delta }^2}+\frac{g_{W} g_{Y} \mu _1^2}{1920 \pi ^2 m_{\Delta }^4}$  \\
	\hline
	$Q_{\text{ll}}$  &  $\frac{3 \text{y$_{\Delta} $}^2 \lambda _{\varphi } m_{\Delta }^2}{128 \pi ^2 m_{\varphi }^4}+\frac{\text{y$_{\Delta} $}^2}{4 m_{\Delta }^2}$  \\
	\hline
	$Q_W$  &  $\frac{g_{W}^3}{1152 \pi ^2 m_{\Delta }^2}$  \\
	\hline
\end{tabular}
}
\end{table*}


	\item{\underline{  $SU(3)_C$ Colored and Complex $SU(2)_L$ doublet ($\tilde{Q}_{3L}$) and singlet ($\tilde{t}_{R}$) model}}
{\small	
	\begin{align}
	\label{supersy}
	\mathcal{L}_{BSM} = \ & \ \mathcal{L}_{SM} + \tilde{Q}_{3L}^{\dagger} (\tilde{k} \tilde{H} \tilde{H}^{\dagger} + k H H^{\dagger} + \lambda_{L} |H|^{2})  \tilde{Q}_{3L} + X_{t} \tilde{Q}_{3L}^{\dagger} \tilde{H} \tilde{t}_{R} + X_{t} \tilde{t}_{R}^{\dagger} \tilde{H}^{\dagger}  \tilde{Q}_{3L} +  \lambda_{R} \tilde{t}_{R}^{\dagger} |H|^{2} \tilde{t}_{R}.
	\end{align}
}
	Here, the heavy fields $\tilde{Q}_{3L}$ and $\tilde{t}_{R}$ contain color charges. The details about these fields are given in Table~\ref{allbsmfields}. Once $\tilde{Q}_{3L}$ and $\tilde{t}_{R}$ are integrated out using \mmaInlineCell{Input}{\mmaDef{CoDEx}}, the effective operators upto dimension-6 for both \mmaInlineCell{Code}{"SILH"} and \mmaInlineCell{Code}{"Warsaw"} bases are generated which are listed in Table~\ref{tab:SUSY}.

\begin{table*}[!hbt]
	\caption{Effective operators and Wilson Coefficients for colored Complex $SU(2)_L$ doublet and singlets models. These two fields are degenerate in mass, i.e., $m_{\tilde{Q}_{3L}}=m_{\tilde{t}_{R}}=m$.}
	\label{tab:SUSY}
	\small
	\centering
	\renewcommand{\arraystretch}{2}
	\subfloat[``SILH'' basis]{
		\begin{tabular}{|*{2}{c|}}
	\hline
	$O_{2B}$  &  $\frac{g_{Y}^2}{320 \pi ^2 m^2}$  \\
	\hline
	$O_{2G}$  &  $\frac{g_{S}^2}{320 \pi ^2 m^2}$  \\
	\hline
	$O_{2W}$  &  $\frac{g_{W}^2}{320 \pi ^2 m^2}$  \\
	\hline
	$O_{3G}$  &  $\frac{g_{S}^2}{320 \pi ^2 m^2}$  \\
	\hline
	$O_{3W}$  &  $\frac{g_{W}^2}{320 \pi ^2 m^2}$  \\
	\hline
	$O_6$  &  $\frac{\tilde{k} \lambda _L X_t^2}{16 \pi ^2 m^4}+\frac{3 \tilde{k} \lambda _L^2}{32 \pi ^2 m^2}+\frac{3 \tilde{k}^2 \lambda _L}{32 \pi ^2 m^2}  + \frac{\tilde{k} X_t^4}{64 \pi ^2 m^6}+\frac{\tilde{k} \lambda _R X_t^2}{32 \pi ^2 m^4} $ \\
	& $ +\frac{\tilde{k}^2 X_t^2}{32 \pi ^2 m^4}+\frac{\tilde{k}^3}{32 \pi ^2 m^2}+\frac{3 k^3}{32 \pi ^2 m^2}+\frac{9 k^2 \lambda _L}{32 \pi ^2 m^2}+\frac{9 k \lambda _L^2}{32 \pi ^2 m^2} $ \\
	& $ +\frac{\lambda _L X_t^4}{64 \pi ^2 m^6}  +\frac{\lambda _L \lambda _R X_t^2}{32 \pi ^2 m^4}  +\frac{\lambda _L^2 X_t^2}{32 \pi ^2 m^4}+\frac{\lambda _L^3}{8 \pi ^2 m^2}+\frac{X_t^6}{320 \pi ^2 m^8} $ \\
	& $ +\frac{\lambda _R X_t^4}{64 \pi ^2 m^6}+\frac{\lambda _R^2 X_t^2}{32 \pi ^2 m^4}+\frac{\lambda _R^3}{16 \pi ^2 m^2}$  \\
	\hline
	$O_{\text{BB}}$  &  $-\frac{\tilde{k}}{2304 \pi ^2 m^2}-\frac{k}{2304 \pi ^2 m^2}-\frac{\lambda _L}{1152 \pi ^2 m^2} $ \\ 
	& $ -\frac{67 X_t^2}{69120 \pi ^2 m^4}-\frac{\lambda _R}{144 \pi ^2 m^2}$  \\
	\hline
	$O_D$  &  $\frac{X_t^2}{320 \pi ^2 m^4}$  \\
	\hline
	$O_{\text{GG}}$  &  $-\frac{\tilde{k}}{192 \pi ^2 m^2}-\frac{\lambda _L}{192 \pi ^2 m^2}-\frac{X_t^2}{384 \pi ^2 m^4}-\frac{\lambda _R}{384 \pi ^2 m^2}$  \\
	\hline
	$O_H$  &  $\frac{\tilde{k} \lambda _L}{8 \pi ^2 m^2}+\frac{3 \tilde{k} X_t^2}{128 \pi ^2 m^4}+\frac{\tilde{k}^2}{64 \pi ^2 m^2}+\frac{3 k \tilde{k}}{32 \pi ^2 m^2} $ \\
	& $  +\frac{k^2}{64 \pi ^2 m^2} +\frac{k \lambda _L}{8 \pi ^2 m^2}+\frac{5 k X_t^2}{128 \pi ^2 m^4} +\frac{\lambda _L X_t^2}{16 \pi ^2 m^4} $ \\
	& $  +\frac{\lambda _L^2}{8 \pi ^2 m^2}+\frac{7 X_t^4}{640 \pi ^2 m^6}+\frac{\lambda _R X_t^2}{16 \pi ^2 m^4} +\frac{\lambda _R^2}{16 \pi ^2 m^2}$  \\
	\hline
	$O_R$  &  $\frac{\tilde{k} X_t^2}{32 \pi ^2 m^4}+\frac{\tilde{k}^2}{32 \pi ^2 m^2}-\frac{k \tilde{k}}{16 \pi ^2 m^2}+\frac{k^2}{32 \pi ^2 m^2} $ \\
	& $  +\frac{k X_t^2}{64 \pi ^2 m^4}+\frac{3 \lambda _L X_t^2}{64 \pi ^2 m^4}+\frac{3 X_t^4}{320 \pi ^2 m^6}+\frac{\lambda _R X_t^2}{32 \pi ^2 m^4}$  \\
	\hline
	$O_T$  &  $-\frac{\tilde{k} X_t^2}{128 \pi ^2 m^4}+\frac{\tilde{k}^2}{64 \pi ^2 m^2}+\frac{k \tilde{k}}{32 \pi ^2 m^2} $ \\
	& $  +\frac{k^2}{64 \pi ^2 m^2}+\frac{k X_t^2}{128 \pi ^2 m^4}+\frac{X_t^4}{640 \pi ^2 m^6}$  \\
	\hline
	$O_{\text{WB}}$  &  $\frac{\tilde{k}}{384 \pi ^2 m^2}-\frac{k}{384 \pi ^2 m^2}+\frac{11 X_t^2}{11520 \pi ^2 m^4}$  \\
	\hline
	$O_{\text{WW}}$  &  $-\frac{\tilde{k}}{256 \pi ^2 m^2}-\frac{k}{256 \pi ^2 m^2}-\frac{\lambda _L}{128 \pi ^2 m^2}-\frac{X_t^2}{2560 \pi ^2 m^4}$  \\
	\hline
	$O_{\text{W}}$  &  $\frac{X_{t}^2}{640 \pi^{2} m^{4} }$  \\	
	\hline
	$O_{\text{B}}$  &  $\frac{X_{t}^2}{640 \pi^{2} m^{4} }$  \\	
	\hline	
\end{tabular}
}
	\subfloat[``Warsaw'' basis]{
\begin{tabular}{|*{2}{c|}}
	\hline
	$Q_G$  &  $\frac{g_S^3}{1920 \pi ^2 m^2}$  \\
	\hline
	$Q_H$  &  $\frac{\tilde{k} \lambda _L X_t^2}{16 \pi ^2 m^4}+\frac{3 \tilde{k} \lambda _L^2}{32 \pi ^2 m^2}+\frac{3 \tilde{k}^2 \lambda _L}{32 \pi ^2 m^2}+\frac{\tilde{k} X_t^4}{64 \pi ^2 m^6} $ \\ 
	 & $ +\frac{\tilde{k} \lambda _R X_t^2}{32 \pi ^2 m^4}+\frac{\tilde{k}^2 X_t^2}{32 \pi ^2 m^4} +\frac{\tilde{k}^3}{32 \pi ^2 m^2}$ \\
	 & $ +\frac{3 k^3}{32 \pi ^2 m^2}+\frac{9 k^2 \lambda _L}{32 \pi ^2 m^2}+\frac{9 k \lambda _L^2}{32 \pi ^2 m^2} $ \\
	 & $  +\frac{\lambda _L X_t^4}{64 \pi ^2 m^6}+\frac{\lambda _L \lambda _R X_t^2}{32 \pi ^2 m^4}+\frac{\lambda _L^2 X_t^2}{32 \pi ^2 m^4}+\frac{\lambda _L^3}{8 \pi ^2 m^2}+\frac{X_t^6}{320 \pi ^2 m^8} $ \\
	 & $  +\frac{\lambda _R X_t^4}{64 \pi ^2 m^6}+\frac{\lambda _R^2 X_t^2}{32 \pi ^2 m^4}+\frac{\lambda _R^3}{16 \pi ^2 m^2}$  \\
	\hline
	$Q_{\text{HB}}$  &  $-\frac{\tilde{k} g_Y^2}{2304 \pi ^2 m^2}-\frac{k g_Y^2}{2304 \pi ^2 m^2}-\frac{g_Y^2 \lambda _L}{1152 \pi ^2 m^2} $ \\ 
	 & $ -\frac{67 g_Y^2 X_t^2}{69120 \pi ^2 m^4}-\frac{g_Y^2 \lambda _R}{144 \pi ^2 m^2}$  \\
	\hline
	$Q_{H\square }$  &  $\frac{\tilde{k} X_t^2}{64 \pi ^2 m^4}+\frac{\tilde{k}^2}{64 \pi ^2 m^2}-\frac{k \tilde{k}}{32 \pi ^2 m^2}+\frac{k^2}{64 \pi ^2 m^2} $ \\
	 & $  +\frac{k X_t^2}{128 \pi ^2 m^4}+\frac{3 \lambda _L X_t^2}{128 \pi ^2 m^4}+\frac{3 X_t^4}{640 \pi ^2 m^6}+\frac{\lambda _R X_t^2}{64 \pi ^2 m^4}$  \\
	\hline
	$Q_{\text{HD}}$  &  $\frac{\tilde{k} \lambda _L}{8 \pi ^2 m^2}+\frac{5 \tilde{k} X_t^2}{64 \pi ^2 m^4}+\frac{\tilde{k}^2}{16 \pi ^2 m^2}+\frac{3 k^2}{16 \pi ^2 m^2}+\frac{3 k \lambda _L}{8 \pi ^2 m^2} $ \\
	 & $  +\frac{5 \lambda _L X_t^2}{64 \pi ^2 m^4}+\frac{\lambda _L^2}{4 \pi ^2 m^2}+\frac{X_t^4}{32 \pi ^2 m^6}+\frac{5 \lambda _R X_t^2}{64 \pi ^2 m^4}+\frac{\lambda _R^2}{8 \pi ^2 m^2}$  \\
	\hline
	$Q_{\text{HG}}$  &  $-\frac{\tilde{k} g_S^2}{192 \pi ^2 m^2}-\frac{g_S^2 \lambda _L}{192 \pi ^2 m^2}-\frac{g_S^2 X_t^2}{384 \pi ^2 m^4}-\frac{g_S^2 \lambda _R}{384 \pi ^2 m^2}$  \\
	\hline
	$Q_{\text{HW}}$  &  $-\frac{\tilde{k} g_W^2}{256 \pi ^2 m^2}-\frac{k g_W^2}{256 \pi ^2 m^2}-\frac{g_W^2 \lambda _L}{128 \pi ^2 m^2}-\frac{g_W^2 X_t^2}{2560 \pi ^2 m^4}$  \\
	\hline
	$Q_{\text{HWB}}$  &  $\frac{\tilde{k} g_W g_Y}{192 \pi ^2 m^2}-\frac{k g_W g_Y}{192 \pi ^2 m^2}+\frac{11 g_W g_Y X_t^2}{5760 \pi ^2 m^4}$  \\
	\hline
	$Q_W$  &  $\frac{g_W^3}{1920 \pi ^2 m^2}$  \\
	\hline
\end{tabular}
}
\end{table*}

\end{enumerate}

\newpage

\subsection{Dimension-6 operators in Warsaw  basis}

Here we have listed all 59 operators in \mmaInlineCell{Code}{"Warsaw"} basis in Table~\ref{Warsaw}.

\begin{table*}[!hbt]
	\caption{Dimension-6 operators in ``Warsaw'' basis.}
	\label{Warsaw}
	\centering
	\small
	\renewcommand{\arraystretch}{2}
	\resizebox{\columnwidth}{!}{%
		\begin{tabular}{|*{6}{c|}}
			\hline
			$\text{}$  &  {\bf Scalar}  &  $Q_{\text{lq}}{}^{(1)}$  &  $\left(\bar{l} \gamma _{\mu }\it{ l)(}\bar{q} \gamma ^{\mu }\it{q} \text{ )}\right.$  &  $\text{}$  &  {\bf Scalar-Fermion}  \\
			\hline
			$Q_H$  &  $\left(H^{\dagger }H )^3\right.$  &  $Q_{\text{lq}}{}^{(3)}$  &  $\left(\bar{l} \tau^a\gamma _{\mu } \it{l} \right) \left(\bar{q} \tau^a\gamma ^{\mu }\it{q} \text{ )}\right.$  &  $Q_{\text{eH}}$  &  $\left(  H^{\dagger }H \right) \text{(}\bar{l}\text{ e }H \text{)+h.c.}.$  \\
			\hline
			$Q_{H\square }$  &  $\left(H^{\dagger }H \text{)$\square $(}H^{\dagger }H \right)$  &  $Q_{\text{ee}}$  &  $\left(\bar{e} \gamma ^{\mu }\it{e} \text{ )(}\bar{e} \gamma _{\mu }\it{e} \text{ )}\right.$  &  $Q_{\text{uH}}$  &  $\left(H^{\dagger }H \right) \text{(}\bar{q} \ \it{ u } \ \tilde{H}\text{)+h.c.}.$  \\
			\hline
			$Q_{\text{HD}}$  &  $\left(H^{\dagger }\mathcal{D}_{\mu }H )^*\right(H^{\dagger }\mathcal{D}^{\mu }H )$  &  $Q_{\text{uu}}$  &  $\left(\bar{u} \gamma ^{\mu }\it{u} \text{ )(}\bar{u} \gamma _{\mu }\it{u} \text{ )}\right.$  &  $Q_{\text{dH}}$  &  $\left(H^{\dagger }H \text{)(}\bar{q} \ \it{d} \ H \text{)+h.c.}\right.$  \\
			\hline
			$\text{}$  &  {\bf Gauge Boson}  &  $Q_{\text{dd}}$  &  $\left(\bar{d} \gamma ^{\mu }\it{d} \text{ )(}\bar{d} \gamma _{\mu }\it{d} \text{ )}\right.$  &  $Q_{\text{Hl}}{}^{(1)}$  &  $\left(H^{\dagger }\it{ i }\overleftrightarrow{\mathcal{D}   }_{\mu }\it{ H }\right) \text{(}\bar{l} \gamma ^{\mu }\it{l} \text{ )}$  \\
			\hline
			$Q_G$  &  $f^{\text{abc}}G_{\rho }{}^{a,\mu }G_{\mu }{}^{b,\nu }G_{\nu }{}^{c,\rho }$  &  $Q_{\text{eu}}$  &  $\left(\bar{e} \gamma ^{\mu }\it{e} \text{ )(}\bar{u} \gamma _{\mu }\it{u} \text{ )}\right.$  &  $Q_{\text{Hl}}{}^{(3)}$  &  $\left(H^{\dagger }\it{ i }\tau ^a \overleftrightarrow{\mathcal{D}   }_{\mu }\it{ H }\right) \text{(}\bar{l} \tau ^a \gamma ^{\mu }\it{l} \text{ )}.$  \\
			\hline
			$Q_{\tilde{G}}$  &  $f^{\text{abc}}\tilde{G}_{\rho }{}^{a,\mu }G_{\mu }{}^{b,\nu }G_{\nu }{}^{c,\rho }$  &  $Q_{\text{ed}}$  &  $\left(\bar{e} \gamma ^{\mu }\it{e} \text{ )(}\bar{d} \gamma _{\mu }\it{d} \text{ )}\right.$  &  $Q_{\text{He}}$  &  $\left(H^{\dagger }\it{ i }\overleftrightarrow{\mathcal{D}   }_{\mu }\it{ H }\right) \text{(}\bar{e} \gamma ^{\mu }\it{e} \text{ )}.$  \\
			\hline
			$Q_W$  &  $\epsilon ^{\text{abc}}W_{\rho }{}^{a,\mu }W_{\mu }{}^{b,\nu }W_{\nu }{}^{c,\rho }$  &  $Q_{\text{ud}}{}^{(1)}$  &  $\left(\bar{u} \gamma^{\mu }\it{u} \text{ )(}\bar{d} \gamma _{\mu }\it{d} \text{ )}\right.$  &  $Q_{\text{Hq}}{}^{(1)}$  &  $\left(H^{\dagger }\it{ i }\overleftrightarrow{\mathcal{D}   }_{\mu }\it{ H }\right) \text{(}\bar{q} \gamma ^{\mu }\it{q} \text{ )}.$  \\
			\hline
			$Q_{\tilde{W}}$  &  $\epsilon ^{\text{abc}}\tilde{W}_{\rho }{}^{a,\mu }W_{\mu }{}^{b,\nu }W_{\nu }{}^{c,\rho }$  &  $Q_{\text{ud}}{}^{(8)}$  &  $\left(\bar{u} \lambda ^a\gamma^{\mu }\it{u} \text{ )(}\bar{d} \lambda ^a\gamma _{\mu }\it{d} \text{ )}\right.$  &  $Q_{\text{Hq}}{}^{(3)}$  &  $\left(H^{\dagger }\it{ i }\tau^a\overleftrightarrow{\mathcal{D}   }_{\mu }\it{ H }\right) \text{(}\bar{q} \tau^a \gamma ^{\mu }\it{q} \text{ )}.$  \\
			\hline
			$\text{}$  &  {\bf Scalar-Gauge Boson}  &  $Q_{\text{le}}$  &  $\left(\bar{l} \gamma^{\mu } l \right) \left(\bar{e} \gamma _{\mu }\it{e} \text{ )}\right.$  &  $Q_{\text{Hu}}$  &  $\left(H^{\dagger }\it{ i }\overleftrightarrow{\mathcal{D}   }_{\mu }\it{ H }\right) \text{(}\bar{u} \gamma ^{\mu }\it{u} \text{ )}$  \\
			\hline
			$Q_{\text{HG}}$  &  $\left(H^{\dagger }H \right)G_{\mu \nu }{}^aG^{a,\mu \nu }$  &  $Q_{\text{lu}}$  &  $\left(\bar{l} \gamma^{\mu } \it{l} \right) \left(\bar{u} \gamma _{\mu }\it{u} \text{ )}\right.$  &  $Q_{\text{Hd}}$  &  $\left(H^{\dagger }\it{ i }\overleftrightarrow{\mathcal{D}   }_{\mu }\it{ H }\right) \text{(}\bar{d} \gamma ^{\mu }\it{d} \text{ )}.$  \\
			\hline
			$Q_{H\tilde{G}}$  &  $\left(H^{\dagger }H \right) \tilde{G}_{\mu \nu }{}^aG^{a,\mu \nu } $  &  $Q_{\text{ld}}$  &  $\left(\bar{l} \gamma^{\mu } \it{l} \right) \left(\bar{d} \gamma _{\mu }\it{d} \text{ )}\right.$  &  $Q_{\text{Hud}}\text{ }$  &  $\left(\tilde{H}^{\dagger }\it{ i }\overleftrightarrow{\mathcal{D}   }_{\mu }\it{ H } \right) \text{(}\bar{u} \gamma ^{\mu }\it{d} \text{ )} \text{+ h.c.} $  \\
			\hline
			$Q_{\text{HW}}$  &  $\left(H^{\dagger }H \right)W_{\mu \nu }{}^aW^{a,\mu \nu }$  &  $Q_{\text{qe}}$  &  $\left(\bar{q} \gamma ^{\mu }\it{q} \text{ )(}\bar{e} \gamma_{\mu }\it{e} \text{ )}\right.$  &  $\text{}$  &  {\bf Fermion-Scalar-Gauge Boson}  \\
			\hline
			$Q_{H\tilde{W}}$  &  $\left(H^{\dagger }H \right)\tilde{W}_{\mu \nu }{}^aW^{a,\mu \nu }$  &  $Q_{\text{qu}}{}^{(1)}$  &  $\left(\bar{q} \gamma_{\mu }\it{q} \text{ )(}\bar{u} \gamma^{\mu }\it{u} \text{ )}\right.$  &  $Q_{\text{eW}}$  &  $\left(\bar{l} \sigma ^{\mu \nu }\it{e} \text{ )}\tau ^a\it{ H }W_{\mu \nu }{}^a\text{+h.c.}\right.$  \\
			\hline
			$Q_{\text{HB}}$  &  $\left(H^{\dagger }H \right)B_{\mu \nu }B^{\mu \nu }$  &  $Q_{\text{qu}}{}^{(8)}$  &  $\left(\bar{q} \gamma _{\mu }\lambda^a\it{q} \text{ )(}\bar{u} \gamma^{\mu }\lambda^a \it{u} \text{ )}\right.$  &  $Q_{\text{eB}}$  &  $\left(\bar{l} \sigma ^{\mu \nu }\it{e} \text{ )\it{ H }}B_{\mu \nu }\text{+h.c.}\right.$  \\
			\hline
			$Q_{H\tilde{B}}$  &  $\left(H^{\dagger }H \right)\tilde{B}_{\mu \nu }B^{\mu \nu }$  &  $Q_{\text{qd}}{}^{(1)}$  &  $\left(\bar{q} \gamma _{\mu }\it{q} \text{ )(}\bar{d} \gamma^{\mu }\it{d} \text{ )}\right.$  &  $Q_{\text{uG}}$  &  $\left(\bar{q} \sigma ^{\mu \nu } \lambda ^a \it{ u  } \right) \tilde{H} G_{\mu \nu }{}^a\text{+h.c.}.$  \\
			\hline
			$Q_{\text{HWB}}$  &  $\left(H^{\dagger }\tau ^aH \right)W_{\mu \nu }{}^aB^{\mu \nu }$  &  $Q_{\text{qd}}{}^{(8)}$  &  $\left(\bar{q} \lambda^a\gamma ^{\mu }\it{q} \text{ )(}\bar{d} \lambda ^a\gamma _{\mu }\it{d} \text{ )}\right.$  &  $Q_{\text{uW}}$  &  $\left(\bar{q} \sigma ^{\mu \nu } \it{u} \right) \tau ^a \tilde{H} W_{\mu \nu }{}^a\text{+h.c.}$  \\
			\hline
			$Q_{H\tilde{W}B}$  &  $\left(H^{\dagger }\tau ^aH \right)\tilde{W}_{\mu \nu }{}^aB^{\mu \nu }$  &  $Q_{\text{ledq}}$  &  $\left(\bar{l}^j\it{e} \text{ )(}\bar{d} q_j\text{)+h.c.}\right.$  &  $Q_{\text{uB}}$  &  $\left(\bar{q} \sigma ^{\mu \nu }\it{ u } \right) \tilde{H} B_{\mu \nu }\text{+h.c.}.$  \\
			\hline
			$\text{}$  &  {\bf Fermions}  &  $Q_{\text{quqd}}{}^{(1)}$  &  $\left(\bar{q}^j\it{u} \text{ )}\epsilon _{\text{jk}}\right(\bar{q}^k\it{d} \text{ )+h.c.}$  &  $Q_{\text{dG}}$  &  $\left(\bar{q} \sigma ^{\mu \nu } \lambda ^a\it{d} \text{ )\it{ H }}G_{\mu \nu }{}^a\text{+h.c.}\right.$  \\
			\hline
			$Q_{\text{ll}}$  &  $\left(\bar{l} \gamma _{\mu }\it{l}  \right) \left(\bar{l} \gamma ^{\mu }\it{l} \text{ )}\right.$  &  $Q_{\text{quqd}}{}^{(8)}$  &  $\left(\bar{q}^j \lambda ^a\it{u} \text{ )}\epsilon _{\text{jk}}\right(\bar{q}^k \lambda ^a\it{d} \text{ )+h.c.}$  &  $Q_{\text{dW}}$  &  $\left(\bar{q} \sigma ^{\mu \nu }\it{d} \text{ )}\tau ^a\it{\it{ H }}W_{\mu \nu }{}^a\text{+h.c.}\right.$  \\
			\hline
			$Q_{\text{qq}}{}^{(1)}$  &  $\left(\bar{q} \gamma _{\mu }\it{q} \text{ )(}\bar{q} \gamma ^{\mu }\it{q} \text{ )}\right.$  &  $Q_{\text{lequ}}{}^{(1)}$  &  $\left(\bar{l}^j\it{e} \text{ )}\epsilon _{\text{jk}}\right(\bar{q}^k\it{u} \text{ )+h.c.}$  &  $Q_{\text{dB}}$  &  $\left(\bar{q} \sigma ^{\mu \nu }\it{d} \text{ )\it{ H }}B_{\mu \nu }\text{+h.c.}\right.$  \\
			\hline
			$Q_{\text{qq}}{}^{(3)}$  &  $\left(\bar{q} \lambda ^a\gamma _{\mu }\it{q} \text{ )(}\bar{q} \lambda ^a\gamma ^{\mu }\text{ q)}\right.$  &  $Q_{\text{lequ}}{}^{(3)}$  &  $\left(\bar{l}^j \sigma _{\mu \nu }\it{e} \text{ )}\epsilon _{\text{jk}}\right(\bar{q}^k \sigma _{\mu \nu }\it{d} \text{ )+h.c.}$  &  $\text{}$  &  $\text{}$  \\
			\hline
		\end{tabular}%
	}
\end{table*}

\subsection{Dimension-6 operators in ``SILH'' bases}

Here we have provided the operators in \mmaInlineCell{Code}{"SILH"} basis in Table~\ref{SILH}.

\begin{table*}
	\caption{Dimension-6 operators in ``SILH'' basis.}
	\label{SILH}
	\centering
	\small
	\renewcommand{\arraystretch}{2}
	\begin{tabular}{|*{4}{c|}}
		\hline
		$O_{\text{GG}}$  &  $g_S{}^2\left(H^{\dagger }H \right)G_{\mu \nu }{}^aG^{a,\mu \nu }$  &  $O_H$  &  $\frac{1}{2}\left(\partial _{\mu }\right(H^{\dagger }H ))^2$  \\
		\hline
		$O_{\text{WW}}$  &  $g_W{}^2\left(H^{\dagger }H \right)W_{\mu \nu }{}^aW^{a,\mu \nu }$  &  $O_T$  &  $\frac{1}{2}\left(H^{\dagger }\overleftrightarrow{\mathcal{D}}_{\mu }H \right)^2$  \\
		\hline
		$O_{\text{BB}}$  &  $g_Y{}^2\left(H^{\dagger }H \right)B_{\mu \nu }B^{\mu \nu }$  &  $O_R$  &  $\left(H^{\dagger }H  \right) \left(\mathcal{D}_{\mu }H )^{\dagger }\right(\mathcal{D}^{\mu }H )$  \\
		\hline
		$O_{\text{WB}}$  &  $2g_Wg_Y\left(H^{\dagger }\tau ^aH  \right) \left(W_{\mu \nu }{}^aB^{\mu \nu }\right)$  &  $O_D$  &  $\left(\mathcal{D}^2H )^{\dagger }\right(\mathcal{D}^2H )$  \\
		\hline
		$O_W$  &  $i g_W \left( H^{\dagger }\tau ^a\overleftrightarrow{\mathcal{D}  }^{\mu }H  \right) \left(\mathcal{D}^{\nu }W_{\mu \nu }{}^a\right)$  &  $O_6$  &  $\left(H^{\dagger }H )^3\right.$  \\
		\hline
		$O_B$  &  $\frac{i}{2}g_Y\left(H^{\dagger }\overleftrightarrow{\mathcal{D}  }^{\mu }H  \right) \left(\partial ^{\nu }B_{\mu \nu }\right)$  &  $O_{2G}$  &  $-\frac{1}{2}\left(\mathcal{D}^{\mu }G_{\mu \nu }{}^a)^2\right.$  \\
		\hline
		$O_{3G}$  &  $\frac{g_S}{3!}f^{\text{abc}}G_{\rho }{}^{a,\mu }G_{\mu }{}^{b,\nu }G_{\nu }{}^{c,\rho }$  &  $O_{2W}$  &  $-\frac{1}{2}\left(\mathcal{D}^{\mu }W_{\mu \nu }{}^a)^2\right.$  \\
		\hline
		$O_{3W}$  &  $\frac{g_W}{3!}\epsilon ^{\text{abc}}W_{\rho }{}^{a,\mu }W_{\mu }{}^{b,\nu }W_{\nu }{}^{c,\rho }$  &  $O_{2B}$  &  $-\frac{1}{2}\left(\partial ^{\mu }B_{\mu \nu })^2\right.$  \\
		\hline
	\end{tabular}
\end{table*}

\end{document}